\documentclass[journal]{IEEEtran}
\usepackage{amsmath,latexsym,amssymb,stmaryrd,pifont}
\usepackage{setspace,stackrel,tikz,graphicx}
\usepackage{exscale,relsize,subfig,textcomp,stackrel,setspace,float}
\usepackage{mdframed,pifont}
\usepackage{multicol,xfrac}
\usepackage{enumitem}
\usepackage{multirow, booktabs}
\usepackage{mathtools}
\usepackage{tabularx}
\usepackage[nolist]{acronym}
\usepackage[input-decimal-markers=.]{siunitx}
\usepackage{longtable}
\usepackage{url}
\usepackage{cuted}  
\usepackage[T1]{fontenc}
\usepackage{fancyhdr}
\usepackage{lipsum}
\usepackage{multicol}
\usepackage{graphicx}
\usepackage[utf8]{inputenc}
\pdfoutput=1
\usepackage{algorithmic,algorithm}

\usepackage{float}
\usepackage{amsthm}
\usepackage{array}
\usepackage{colortbl}
\usepackage{xcolor}
\usepackage{mathtools,amssymb,lipsum}

\usepackage{cuted}
\usepackage[most]{tcolorbox}
\begin{acronym}
   \acro{ISAC}{integrated sensing and communications}
   \acro{3GPP}{3rd Generation Partnership Project}
   \acro{TR}{technical report}
   \acro{RCS}{radar cross-section}
   \acro{GBSM}{geometry-based stochastic model}
   \acro{LoS}{line-of-sight}
    \acro{NLoS}{non line-of-sight}
    \acro{AI}{artificial intelligence}
    \acro{MIMO}{multiple-input multiple-output}
    \acro{mmWave}{millimeter-wave}
    \acro{AoA}{azimuth angle of arrival} 
    \acro{AA}{angles of arrival}
     \acro{ZoA}{zenith angle of arrival} 
    \acro{AoD}{azimuth angle of departure} 
     \acro{ZoD}{zenith angle of departure} 
     \acro{AD}{angles of departure}
   \acro{ToA}{time-of-arrival}
   \acro{SMC}{shared multipath component}
   \acro{HAR}{human activity recognition}
   \acro{Rx}{receiver}
   \acro{Tx}{transmitter}
   \acro{UAV}{unmanned aerial vehicle}
   \acro{IAB}{integrated access and backhaul}
   \acro{EO}{environmental object}
   \acro{GR}{ground reflection}
   \acro{TRP}{transmission reception point}
   \acro{RF}{radio-frequency}
   \acro{UE}{user equipment}
   \acro{BS}{base station}
   \acro{SP}{scattering point}
    \acro{LSP}{large scale parameters}
      \acro{SSP}{small scale parameters}
   \acro{DS}{delay spread}
   \acro{AS}{angular spread}
   \acro{SF}{shadow fading}
\acro{ASD}{azimuth spread of departure}
\acro{ZSD}{zenith spread of departure}
\acro{ASA}{azimuth spread of arrival}
\acro{ZSA}{zenith spread of arrival}
\acro{XPR}{cross polarization ratio}
\acro{CIR}{channel impulse response}
\acro{BER}{bit-error rate}
\acro{UMa}{Urban Macro}
\acro{UMi}{Urban Micro}
\acro{RMa}{Rural Macro}
\acro{InH}{Indoor Hotspot}
\acro{InF}{Indoor Factory}
\acro{InF-DH}{Indoor Factory with Dense clutter and High base station height}
\acro{ROC}{receiver operating characteristic}
\acro{OFDM}{Orthogonal frequency-division multiplexing}
\acro{QAM}{Quadrature amplitude modulation}
\acro{CP}{cyclic prefix}
\acro{URA}{uniform rectangular array}
\acro{SnC}[S\&C]{sensing and communication}
\acro{NR}{New Radio}
\acro{USRP}{universal software-peripheral radio}
\acro{ZC}{Zadoff-Chu}
\acro{RFIC}{radio-frequency integrated circuit}
\acro{IF}{intermediate frequency}
\acro{SISO}{single-input single-output}
\acro{SVD}{singular value decomposition}
\acro{SNR}{signal-to-noise ratio}
\acro{ED}{energy detector}
\acro{CPM}{cross-polarization matrix }
\acro{CM}{channel modeling}
\acro{SCM}{spatial channel model}
\acro{UMTS}{universal mobile telecommunications system}
\acro{CC}{channel coefficients}
\acro{CG}{channel gain}
\end{acronym}

\usepackage{xcolor,cite,etoolbox}
\makeatletter 
\pretocmd\@bibitem{\color{black}\csname keycolor#1\endcsname}{}{\fail}
\newcommand\citecolor[1]{\@namedef{keycolor#1}{\color{blue}}}
\makeatother

\usepackage{etoolbox} 

\usepackage{titlesec}

\titlespacing{\subsection}{1pt}{*0.05}{*0.1}
\titlespacing{\subsubsection}{3pt}{*0.05}{*0.1}
\begin{document}

\title{\vspace{0cm}\LARGE 
A Framework for Geometry-based Statistical Channel Modeling in ISAC Systems\vspace{0em}}
\makeatletter
\patchcmd{\@maketitle}
  {\addvspace{0\baselineskip}\egroup}
  {\addvspace{0\baselineskip}\egroup}
  {}
  {}
\makeatother
\author{Ali Waqar Azim, Ahmad Bazzi, Theodore S. Rappaport, Marwa Chafii
\thanks{This work is supported by Tamkeen under the Research Institute NYUAD grant CG017, and NYUAD Center for Artificial Intelligence and Robotics, funded by Tamkeen under the Research Institute Award CG010.\\Ali Waqar Azim is with James Watt School of Engineering, University of Glasgow, G12 8QQ, UK. Ahmad Bazzi and Marwa Chafii are with the Engineering Division, New York University (NYU) Abu Dhabi, 129188, UAE
(email: aliwaqarazim@gmail.com, \{ahmad.bazzi,marwa.chafii\}@nyu.edu). Ahmad Bazzi, Theodore S. Rappaport, and Marwa Chafii are with NYU WIRELESS, NYU Tandon School of Engineering, Brooklyn, 11201, NY, USA .}}
\maketitle

\renewcommand{\headrulewidth}{0pt}

\begin{abstract}
 This paper proposes a comprehensive framework for a {geometry-based statistical model for} integrated sensing and communication (ISAC) tailored for bistatic systems. Our dual-component model decomposes the ISAC channel into a target channel encompassing all multipath components produced by a sensing target {parameterized} by the target's radar cross-section and scattering points, and a background channel comprising all other propagation paths that do not interact with the sensing target. The framework extends TR38.901 via a hybrid clustering approach, integrating spatiotemporally consistent deterministic clusters with stochastic clusters to preserve channel reciprocity and absolute delay alignment for sensing parameter estimation. Extensive simulations across {urban macro, urban micro, and indoor factory} scenarios demonstrate that the model maintains communication performance parity with the standard TR38.901, validated through bit-error rate analysis obtained via simulated and measured ISAC channels and channel capacity assessment, while enabling sensing performance evaluation, such as target ranging error for localization and receiver operating characteristic curves for detection probability.
\end{abstract}
\IEEEpeerreviewmaketitle
\vspace{-0.1cm}
\section{Introduction}
\Ac{ISAC} is a transformative paradigm for 6G systems, enabling joint \ac{SnC} functionalities through shared resources \cite{liu2023beginning}. This resource sharing requires a unified channel model that supports reliable communication and accurate sensing. Unlike conventional communication channels, the \ac{ISAC} channel must jointly model two components: (1) the \emph{background channel} representing propagation independent of sensing targets, and (2) the \emph{target channel} representing reflections from sensing targets \cite{yang2024integrated}. Target-induced returns manifest as additional resolvable paths in the \ac{CIR}, with time-varying complex amplitudes and associated Doppler shifts.

While modern industry standards like the \ac{3GPP} provide frameworks for channel modeling, their parameters are rooted in foundational early research on \ac{GBSM}. Early work in \cite{501430} and \cite{587631} established the geometric relationship between scatterer placement and multipath characteristics, which bridges the gap between simplified stochastic models and site-specific ray-tracing, especially in multipath environments \cite{501477,liberti1999smart}. The works shifted channel modeling from purely stochastic methods to those incorporating the physical geometry of scatterers \cite{312781,501430,liberti1999smart}. More recently, \cite{7248689} developed a 3D statistical model based on extensive NYC measurements, similar in spirit to \ac{GBSM} but adapted for mmWave frequencies. In addition, \cite{9411894} extends spatial geometric concepts to indoor environments for 5G and 6G system design and \cite{7999294} compares various modeling approaches, including geometric and stochastic hybrids.

Current standardized legacy \ac{GBSM} channel models, such as \ac{3GPP} \ac{TR}38.901 \cite{tr}, are limited for \ac{ISAC} applications. These models characterize propagation through statistical clusters and stochastically defined interaction processes, thereby lacking explicit, geometrically consistent scatterer representation needed for sensing. Furthermore, this statistical approach also neglects critical target-specific features and phenomenology, such as \ac{RCS} and associated Doppler spreads. Although site-specific ray-tracing can implicitly capture target-channel interactions, its prohibitive computational complexity precludes large-scale system simulation. Recognizing this gap, \ac{3GPP} has introduced new study items in Release 19 dedicated to \ac{ISAC} \ac{CM} \cite{3gpp_R1-2504945}. The resulting \ac{CM} framework must therefore be architected around a unified representation of the target and background channels. This requires a shift from purely stochastic cluster modeling to a hybrid approach where \acp{EO} are represented deterministically, defined by parameters such as 3D position, orientation, and velocity, while inherently supporting spatiotemporal consistency, extended object representation, and channel reciprocity \cite{wymeersch2025cross}. Here, spatiotemporal consistency is defined as the property where scattering clusters possess pre-defined locations and deterministic reflection properties. 
\subsection{{Evolution of \ac{3GPP} Geometric Stochastic Channel Models}}
{The development of \ac{TR}38.901 \cite{tr} results from steady advancements in communication systems, where the progression began with the \ac{SCM} in \ac{TR}25.996 for 3G/UMTS, which introduced a geometric stochastic approach limited to 2D spatial propagation in the azimuth plane \cite{3GPP_TR_25_996}. LTE systems further necessitated elevation modeling, leading to the 3D channel model in \ac{TR}36.873 \cite{3GPP_TR_36_873}, which then extended the \ac{SCM} framework to 3D space by incorporating zenith \ac{AA} and \ac{AD}, creating a full 3D spatial channel framework. \ac{TR}38.901 \cite{tr} is the consolidation of these prior models. It inherits stochastic frameworks from \ac{TR}36.873 and extends it to support higher frequency bands (up to \SI{100}{\GHz}), a wider range of deployment scenarios, and more detailed parameterization. However, despite its sophistication for communication-centric simulations, its core approach remains based on the statistical characterization of propagation clusters, which poses inherent limitations for \ac{ISAC} and as such, cannot be used for \ac{ISAC} system evaluation.}
The proposed sensing channel representation can also support network-level decisions in emerging \ac{IAB} architectures. More specifically, the network can identify locations that support high-throughput, relatively unobstructed wireless backhaul links at millimeter wave bands, and distinguish them from regions where the channel is poorly suited for wireless backhaul and alternative (e.g., wired or out-of-band) solutions are preferable, which creates a provisioning mechanism whereby the same \ac{ISAC} framework used for radio environment sensing can guide \ac{IAB} planning and dynamic backhaul selection, complementing early work on  \ac{IAB} at millimeter wave frequencies \cite{6666553}.
\subsection{Related Work}
Recent advances in \ac{ISAC} systems have spurred significant progress in \ac{ISAC} \ac{CM}, addressing its unique challenges. {Different scatterer-based hybrid models have been proposed, combining deterministic and stochastic components for improved accuracy, such as \cite{chen2023scatterer}.} 
For bistatic \ac{ISAC} systems, recent frameworks introduce weak-power cluster retention for target characterization \cite{luo2024channel}. Furthermore, cluster-based \ac{ISAC} \ac{CM} framework in \cite{ye2024general} decomposes the environment into \ac{LoS}, \ac{NLoS}, and clutter components, whose weighted summation yields the effective \ac{CIR}, effectively integrating sensing characteristics into communication models. \cite{liu2024extend} introduces a \ac{3GPP}-compliant \ac{GBSM} \cite{501430,587631} to support \ac{ISAC} by introducing sharing feature that enables flexible modeling of shared clusters, scatterers, and propagation paths across communication and sensing channels within a unified stochastic framework. \cite{zhang2023shared} presents a shared multipath component evolution model to characterize the correlation between communication and sensing channels in \ac{ISAC}, validated through \SI{28}{\GHz} measurements. Hybrid methodologies combining statistical and deterministic techniques have also emerged, leveraging multi-scattering center models for precise target characterization \cite{chen2024multi}. Beyond physical-layer modeling, recent surveys highlight features such as \ac{RCS} integration and \ac{AI}-enhanced frameworks for sub-\SI{6}{\GHz}/mmWave bands \cite{zhang2023integrated}. In this context, NYURay \cite{kanhere2019map}, a site-specific 3D ray tracer developed at NYU WIRELESS, can be used for \ac{ISAC} \cite{bazzi2025isac} owing to its site-specific ray tracing with the very same angular and temporal parameters a sensing algorithm will use, while NYUSIM \cite{poddar2023tutorial} can provide statistical insights because of the stochastic channel and can be useful for large-scale simulation studies requiring system- or link-level testing, beamforming evaluation, and statistical coverage/capacity analysis.
Recent work has demonstrated that NYURay can be pushed to the mobile or edge to perform multi-stage location optimization by aligning measured and simulated power-delay profiles and to calibrate and validate the NYURay simulator at FR3 frequencies \cite{ying2025multistage,ying2025sitespecific}.
\subsection{Novelty and Contributions}
\begin{table*}[t]
\caption{Comparison of the proposed \ac{ISAC} channel-modeling framework with recent related works.}
\label{tab:comparison_framework}
\centering
\footnotesize
\setlength{\tabcolsep}{4pt}
\renewcommand{\arraystretch}{1.2}
\begin{tabular}{|p{1cm}|p{3.2cm}|p{2.5cm}|p{2.8cm}|p{2.4cm}|p{4.4cm}|}
\hline
\textbf{Work} & \textbf{Core Modeling Idea} & \textbf{Relation to \ac{3GPP} / \ac{TR}38.901} & \textbf{Target/Scatterer Representation} & \textbf{Shared-Cluster / Correlation Modeling} & \textbf{Main Distinction from the Proposed Framework} \\
\hline

\textbf{\cite{luo2024channel}} &
Extends bistatic \ac{ISAC} channel under the \ac{3GPP} framework by retaining weaker clusters and converting selected clusters into sensing clusters according to the sensing scenario. &
Strong \ac{3GPP} compatibility; follows the \ac{3GPP} procedure with added weak-cluster retention. &
Sensing clusters can be modeled statistically or deterministically; emphasis is on target-cluster selection / conversion. &
Not the main focus. &
The proposed framework explicitly decomposes the \ac{ISAC} channel into target and background components, parameterizes the target channel via \ac{RCS} and SPs, supports deterministic and stochastic target placement, and incorporates \acp{EO}. \\
\hline

\textbf{\cite{ye2024general}} &
Constructs an \ac{ISAC} channel by building 3D scattering-cluster positions and combining \ac{LoS}, \ac{NLoS}, and clutter echoes through weighted summation. &
Built in a \ac{3GPP}-compatible manner for communication-channel generation. &
Scatterers / clusters are central, and the effective sensing response is formed from categorized echo components. &
Not the main focus. &
Proposed framework is based on \ac{TR}38.901-preserving hybrid stochastic/deterministic cluster-generation methodology with explicit target/background separation, rather than an echo-category summation model. \\
\hline

\textbf{\cite{liu2024extend}} &
Proposes a \ac{3GPP}-compliant 3D \ac{GBSM} with a sharing feature, where communication and sensing \acp{CIR} are composed of shared and non-shared clusters. &
Explicitly designed to remain compatible with \ac{3GPP} 3D \ac{GBSM}. &
Targets / scatterers are linked to communication clusters through replacement / matching rules. &
Main focus is shared clusters, shared scatterers, and sharing parameter. &
The proposed framework does not take shared-cluster identity as the primary modeling principle; instead, it adopts a dual-component target/background formulation with deterministic geometry-anchored target/\ac{EO} clusters. \\
\hline

\textbf{\cite{zhang2023shared}} &
Develops a \ac{SMC} evolution model to characterize the correlation between communication and sensing channels. &
Not a full \ac{TR} 38.901 channel-generation framework; mainly focused on correlation characterization. &
Multipath components / \acp{SMC} are characterized from measurements; the model is measurement-driven. &
Main focus is communication-sensing correlation via \ac{SMC} evolution probability. &
The proposed framework is a generative \ac{ISAC} channel-synthesis model, not only a correlation model; it provides explicit target/background decomposition, deterministic clusters, and \ac{EO} support for communication and sensing evaluation. \\
\hline

\textbf{Proposed} &
\ac{TR}38.901-structured dual-component \ac{ISAC} \ac{GBSM} with a target channel plus a background channel, combining stochastic generation with deterministic geometry-anchored clusters. &
Keeps the procedure as close as possible to \ac{TR}38.901 while introducing only the extensions required for \ac{ISAC}. &
Explicit \ac{RCS} and \ac{SP}-based target modeling, deterministic and stochastic target placement, and Type-I \ac{EO} modeling. &
Shared-cluster identity is not enforced as the main modeling axis. &
Provides a physically grounded generative framework with absolute-delay consistency, spatiotemporal consistency, reciprocity, and direct support for downstream sensing studies. \\
\hline
\end{tabular}
\end{table*}
To this end, the novelty of the proposed \ac{ISAC} channel model lies in its methodological positioning and integration scope, i.e., we keep the channel-generation procedure as close as possible to the standardized \ac{3GPP} \ac{TR}38.901 fully stochastic pipeline \cite{3gpp38901} while introducing extensions that are essential for \ac{ISAC}. In particular, the background link and the stochastic components of the \ac{Tx}-target and target-\ac{Rx} links follow \ac{TR}38.901 processing, whereas the coefficient-generation pipeline is consistently extended to also accommodate deterministic clusters. This results in a hybrid model in which \ac{NLoS} clusters in both \ac{Tx}-target and target-\ac{Rx} links are explicitly partitioned into stochastic and deterministic subsets, thereby augmenting the \ac{TR}38.901 stochastic cluster set with deterministic, geometry-anchored contributions.

This novelty is also explicit relative to recent prior work. Unlike \cite{luo2024channel}, which extends the \ac{3GPP} bistatic \ac{ISAC} framework mainly by retaining weaker clusters and modeling target clusters in deterministic or statistical form, the proposed framework introduces an explicit dual-component decomposition into target and background channels, with the target channel parameterized by \ac{RCS} and scattering points rather than only by sensing-cluster retention. Unlike \cite{ye2024general}, which constructs the \ac{ISAC} response by categorizing echoes into \ac{LoS}, \ac{NLoS}, and clutter components and combining them through weighted summation after 3D scatterer-position construction, our framework is centered on a \ac{TR}38.901-preserving hybrid stochastic/deterministic cluster-generation mechanism rather than an echo-category summation model. Unlike \cite{liu2024extend}, whose main focus is the sharing feature between communication and sensing channels through shared and non-shared clusters, our framework is centered on explicit target/background decomposition and deterministic EO/target representation within a bistatic generative framework. Unlike \cite{zhang2023shared}, which develops a measurement-based shared multipath component evolution model to characterize communication-sensing correlation at \SI{28}{\GHz}, the proposed work provides a full generative channel-synthesis framework for \ac{ISAC} evaluation across communication and sensing tasks, while remaining compatible with standardized scenario generation. Table \ref{tab:comparison_framework} summarizes the main methodological differences between the proposed framework and recent \ac{ISAC} channel models in \cite{luo2024channel}-\cite{zhang2023shared} highlighting that the proposed work is distinguished by explicit target/background decomposition, \ac{TR}38.901-preserving hybrid stochastic/deterministic generation, and \ac{EO}-aware target modeling.

Beyond this, the framework supports two target placement modes: (i) deterministic placement by specifying the target's 3D Cartesian coordinates and deriving the corresponding delays/angles, and (ii) stochastic placement by selecting a subset of \ac{TR}38.901 generated clusters as target clusters and subsequently estimating their 3D coordinates. Finally, as per \ac{3GPP} suggestion \cite{3gpp_R1-2504945}, we also introduce \acp{EO} in \ac{ISAC} channel modeling framework. In particular, Type-I \acp{EO} are modeled with target-like attributes and are represented through deterministic clusters to enforce spatiotemporal consistency tied to physically meaningful objects. This explicit \ac{EO} taxonomy and \ac{EO}-attribute integration is not modeled in previous studies, which focus on target/scatterer/clutter constructions rather than \ac{EO} entities as defined here.

Against the given background, design insights, and related work, the main contributions of this work are as follows:
\begin{enumerate}[label=\textbf{\arabic*.}, leftmargin=*, labelindent=0.2pt, labelsep=0.5em, nosep]
\item \textbf{A \ac{TR}38.901-structured dual-component \ac{ISAC} \ac{GBSM} framework with \ac{EO} and target-placement support:} Motivated by the \ac{3GPP} recommendations in \cite{3gpp_R1-2504945}, we propose a dual-component \ac{ISAC} channel model comprising a \emph{target channel} and a \emph{background channel}. The target channel explicitly captures the bistatic propagation via a sensing target, parameterized by its \ac{RCS} and \acp{SP}, whereas the background channel accounts for all non-target propagation mechanisms. Moreover, because the target channel is represented through geometrically consistent \acp{SP} and time-varying propagation parameters, the framework can also support downstream sensing tasks such as target classification and human activity recognition. Within a \ac{TR}38.901-structured channel-generation pipeline, the framework further supports both deterministic and stochastic target placement and incorporates \acp{EO} (in particular, Type-I \acp{EO} with target-like attributes) through deterministic clusters. It is emphasized that placing deterministic targets in a \ac{TR}38.901 environmental setup with clutter interaction is one of the fundamental contributions of the paper. The proposed model accommodates all four fundamental bistatic \ac{ISAC} configurations defined by \ac{3GPP} in \cite{3gpp_R1-2504945}, i.e., \ac{TRP}-\ac{TRP}, \ac{TRP}-\ac{UE}, \ac{UE}-\ac{TRP}, and \ac{UE}-\ac{UE} by introducing the corresponding mobility scenarios. Moreover, our framework explicitly provides the channel coefficients for all combinations of \ac{LoS}/\ac{NLoS} propagation states across the \ac{Tx}-target and target-\ac{Rx} links.
\item \textbf{{A hybrid clustering methodology:}} {To bridge the gap between legacy stochastic models and the geometric fidelity required for sensing, we introduce a novel hybrid stochastic-deterministic clustering approach. Our model augments the standard stochastic clusters from \ac{TR}38.901 \cite{tr} with geometrically defined deterministic clusters that exhibit strict spatiotemporal consistency and absolute delay alignment. This hybrid framework enables a full \ac{CIR} characterization for all propagation paths including \ac{LoS} and \ac{NLoS} components within both the background channel and the target channel, thereby ensuring critical properties for sensing, such as channel reciprocity.}

\item {\textbf{Unified performance validation demonstrating dual-functionality:} Our proposed model is validated through simulations across standardized \ac{UMa}, \ac{UMi}, and \ac{InF} scenarios and is further corroborated by real-world channel measurements. This validation demonstrates that the proposed framework maintains full compatibility with \ac{TR}38.901 \cite{tr} for communication performance while simultaneously enabling accurate sensing performance evaluation.}
\end{enumerate}
\subsection{Notation}
Unless otherwise mentioned, this paper adheres to the following notational conventions. Scalars are represented by italic letters, e.g., $N$, $\sigma$, $\kappa$, while vectors and matrices are denoted by boldface uppercase and boldface uppercase calligraphic letters, respectively. {The transpose, Euclidean norm, and Hermitian norm are denoted by $(\cdot)^\mathrm{T}$, $\Vert \cdot\Vert$, and $\Vert \cdot\Vert^2_2$, respectively. The magnitude and dot product are expressed as $|\cdot|$ and $(\cdot, \cdot)$.} The imaginary unit is defined as $j \triangleq \sqrt{-1}$, and the Dirac delta function is $\delta(\cdot)$. 
\subsection{Organization}
The remainder of this paper is organized as follows. Section \ref{sec2} details the proposed geometry-based \ac{ISAC} channel model. Section \ref{sec3} presents the comprehensive framework for bistatic \ac{ISAC} \ac{CM}, outlining the procedural steps for generating general parameters, small-scale parameters, and \ac{CC}. Section \ref{sec4} provides performance evaluation, analyzing communication metrics such as \ac{BER} and ergodic channel capacity, and assessing sensing performance through target ranging accuracy and {\ac{ROC}} curves. Finally, Section \ref{sec5} concludes the paper and suggests directions for future research.
\section{Modeling of Components of ISAC Channel}\label{sec2}
As agreed by \ac{3GPP} {\cite{3gpp_R1-2504945}}, the \ac{ISAC} \ac{CC} between \(u\)th \ac{Rx} antenna element and \(s\)th \ac{Tx} antenna element, \(\boldsymbol{\mathcal{H}}^\mathrm{ISAC}_{u,s}(t,\tau)\) consists of: (1) the target channel, \(\boldsymbol{\mathcal{H}}^\mathrm{tar}_{u,s}(t,\tau)\) and (2) the background channel, \(\boldsymbol{\mathcal{H}}^\mathrm{back}_{u,s}(t,\tau)\), where \(t\) is the time snapshot and \(\tau\) is the delay. \(\boldsymbol{\mathcal{H}}^\mathrm{tar}_{u,s}(t,\tau)\) comprises all the channel components {interacting with the target}. In contrast, \(\boldsymbol{\mathcal{H}}^\mathrm{back}_{u,s}(t,\tau)\) accounts for all propagation paths that do not undergo interaction with the target. The \ac{ISAC} \ac{CC} is given as \cite{3gpp_R1-2504945}:
\begin{equation}\label{ISAC_chan}
 \boldsymbol{\mathcal{H}}^\mathrm{ISAC}_{u,s}(t,\tau) = \boldsymbol{\mathcal{H}}^\mathrm{tar}_{u,s}(t,\tau) + \boldsymbol{\mathcal{H}}^\mathrm{back}_{u,s}(t,\tau). 
\end{equation}

The inherent correlation between the background and target channels arises from their co-propagation through a shared physical environment, where signals may interact with a common set of scatterers. 

Note that the decomposition of the \ac{ISAC} channel is useful not only for channel generation but also for sensing processing. When the background channel is known in advance or estimated from calibration measurements without targets, its contribution can be suppressed to isolate the target response. The resulting residual channel is then used for sensing tasks such as range estimation and target discrimination.

A key advantage of the this framework is that both target and background components are generated with physically consistent absolute delays. As a result, after background suppression, the dominant target echoes remain well localized in time, which improves the reliability of target detection and range estimation. In \ac{NLoS} conditions, stochastic clusters act as clutter and may degrade sensing accuracy, whereas deterministic clusters preserve spatial consistency and still support physically meaningful sensing analysis.
\subsection{Shared Clustering and Environment Coupling}
Recent \ac{ISAC} channel studies highlight that \ac{SnC} links can be statistically coupled due to co-propagation through the same environment and potential interaction with common shared scatterers \cite{liu2024extend}. In the proposed framework, the \ac{ISAC} channel is decomposed into a target component and a non-target background component. The background channel is modeled as a \ac{TR}38.901-compliant \ac{Tx}-\ac{Rx} channel consisting of the \ac{LoS} term and stochastic \ac{NLoS} clusters, whereas target channel is generated via a hybrid mechanism in which a subset of \ac{NLoS} clusters is treated deterministically (geometrically anchored), enabling target-/\ac{EO}-dependent physical effects (e.g., geometry/material-driven \ac{XPR}) in the deterministic part.

In this  proposed framework, explicit shared-cluster identity across the background and the target channels is not enforced, and the \ac{LSP} are generated for the background and independently for the \ac{Tx}-target and target-\ac{Rx} links. Nevertheless, the framework is compatible with incorporating shared clustering and environment coupling by (i) defining a subset of background clusters to deterministic objects with fixed 3D support and reusing them across background and target channels, and/or (ii) introducing cross-link correlation in \ac{LSP} generation (e.g., correlated shadowing/spread fields) to match measured \ac{ISAC} coupling levels.
\vspace{-0.1cm}
\subsection{Target Characterization and Channel Modeling}\label{sec2a}
\subsubsection{Target Characterization}
In \ac{ISAC}, a target is an object intended for detection, localization, or tracking. Depending on the application, targets can be vehicles, \acp{UAV}, or other objects of interest. A target is characterized by its \ac{RCS}, denoted by \(\sigma\), which quantifies its scattering behavior and the power reflected towards the sensing \ac{Rx}. Targets are typically modeled as either a \textit{point target} or an \textit{extended target}. An extended target comprises multiple distributed \acp{SP}, resulting in an aggregate return signal that is a superposition of contributions from its individual scatterers, unlike the single dominant reflection from a point target. It is highlighted that within the proposed framework, the \ac{SP}-level \ac{RCS} values are provided as inputs for target characterization. These \ac{RCS} values can be can be drawn from empirical distributions obtained from pre-performed measurements or from system-level studies.

It is important to note that, unlike \ac{TR}38.901 \cite{tr}, in which all clusters are modeled stochastically, we introduce additional deterministic clusters in the target channel. This allows us to model the target as a separate, controllable component in the \ac{ISAC} channel simulations, enabling localization in addition to sensing. In fact, besides only detecting targets, the sensed targets/scatterers become resources for communications, namely \ac{ISAC} sensing identifies the angular positions and relative strengths of key multipath components, which can then be leveraged for power combining over different angular paths, closely related to early work on range extension via multi-path beam combining \cite{6884191}.
\vspace{-0.1cm}
\paragraph{Application Perspective: Human Activity Recognition}
Although the proposed framework is not presented as a dedicated \ac{HAR} model, it can provide a physically grounded channel-generation layer for \ac{HAR}-oriented \ac{ISAC} studies. Specifically, the decomposition \(\boldsymbol{\mathcal{H}}_{\mathrm{ISAC}}(t,\tau)=\boldsymbol{\mathcal{H}}_{\mathrm{back}}(t,\tau)+\boldsymbol{\mathcal{H}}_{\mathrm{tar}}(t,\tau)\) enables a sensing \ac{Rx} to first estimate or calibrate the background channel \(\boldsymbol{\mathcal{H}}_{\mathrm{back}}(t,\tau)\) and then suppress it in order to isolate the target-dependent component \(\boldsymbol{\mathcal{H}}_{\mathrm{tar}}(t,\tau)\). Within this framework, a human can be represented as an extended target composed of multiple time-varying \acp{SP} associated with major body parts, where the delay, angle, Doppler, and power evolution of these \acp{SP} are driven by body motion over time. The resulting target-channel sequence may then be processed in multiple ways. A data-driven machine learning approach can treat isolated channel realizations or their delay-angle-Doppler representations as labeled inputs for activity classification. On the other hand, a parametric approach can estimate the dominant propagation parameters or reconstruct reflection points/images from the channel state information and infer the activity from the temporal evolution of the reconstructed body structure \cite{bazzi2025isac}. More generally, simpler model-based \ac{HAR} can also be performed by tracking activity-dependent changes in \ac{ToA}, \ac{AA}, and Doppler signatures over time. The proposed model provides the channel framework on top of which such downstream \ac{HAR} algorithms can be built.

In the context of \ac{ISAC} channel  modeling, there are multiple ways target(s) can be modeled in the spatial environment:
\paragraph{{Deterministic Target Placement}}
{A target can be dropped in the \ac{Tx}-\ac{Rx} environment by specifying its known 3D Cartesian coordinates \((x,y,z)\). Once the coordinates are ascertained, geometric parameters such as the propagation delay, \ac{AA}, and \ac{AD} are deterministically derived from the spatial geometry of the \ac{Tx}, target, and \ac{Rx}.}
\paragraph{{Stochastic Target Placement}}
{Targets can also be modeled by designating a subset of stochastically generated clusters from \ac{TR}38.901 \cite{tr} framework as target clusters. However, these clusters lack geometric consistency as they are not anchored to specific 3D coordinates. To address this, the target cluster's 3D coordinates must be determined post-generation, typically using approaches as in \cite{jaeckel2014quadriga,luo2024channel,ye2024general}. In Section \ref{sec3b}, we present a method for deriving the target's 3D coordinates.}
\subsubsection{Environmental Object Characterization and Modeling}
\Acp{EO} also need to be considered for the \ac{ISAC} \ac{CM} framework \cite{3gpp_R1-2504945}. \acp{EO} are non-target objects that have known spatial locations and are categorized by their physical and scattering characteristics relative to the sensing target. Unlike the stochastic cluster, \acp{EO} are deterministic positions and scattering properties. Type I \acp{EO} exhibit target-like properties, including a defined \ac{RCS}, velocity profile, and spatially consistent \acp{SP}. In the proposed framework, the \ac{EO} \ac{RCS} values for each \ac{SP} are used as inputs for \ac{EO} characterization. Type II \acp{EO} represent large-scale environmental structures (e.g., buildings, walls) that significantly exceed typical target dimensions and are modeled using the ground-reflection approach in Section 7.6.8 of \ac{TR}38.901 \cite{tr}, incorporating specular reflection components as deterministic contributions under \ac{NLoS} conditions.
\subsubsection{Target Channel Modeling}
\begin{figure}[t]
    \centering
    \includegraphics[trim={1mm 1mm 1mm 1mm},clip,width=0.49\textwidth]{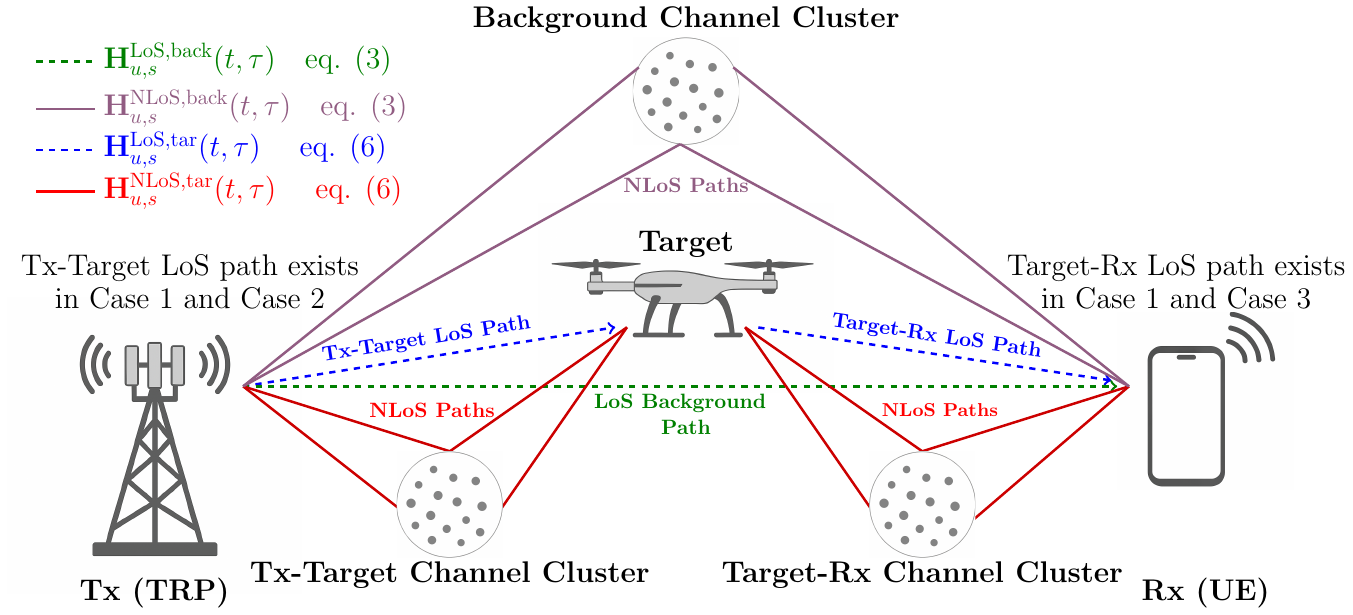} 
    \caption{Propagation conditions for the \ac{ISAC} channel. }
    \label{prop_cond} 
\end{figure}

{The target channel comprises two cascaded links, that are, the \ac{Tx}-target link and the target-\ac{Rx} link. \ac{3GPP} \acp{TR}, such as \ac{TR}38.901 \cite{tr}, \ac{TR}36.777, \ac{TR}37.885 serve as a reference to determine the probability of \ac{LoS} for each link. Furthermore, the target channel \ac{LoS}/\ac{NLoS} condition is determined jointly by both constituent links \cite{3gpp_R1-2504945}. The target channel is \ac{LoS} only when both segments are \ac{LoS}, otherwise, it is \ac{NLoS}, which results in four possible propagation states for the target channel, as shown in Table \ref{propation_conditions} and Fig. \ref{prop_cond}.}
\begin{table}[tb]
    \centering
     \caption{Target Channel Propagation Conditions.}
     \footnotesize
    \begin{tabular}{|c|c|c|}
        \hline
        \rowcolor{gray!20}
        \textbf{Case} & \textbf{Tx-Target} & \textbf{Target-Rx} \\
        \hline
        1 & \ac{LoS}&\ac{LoS} \\
        \hline
        2 & \ac{LoS} &\ac{NLoS} \\
        \hline
        3 & \ac{NLoS} & \ac{LoS} \\
        \hline
        4 & \ac{NLoS} & \ac{NLoS} \\
        \hline
    \end{tabular} 
    \label{propation_conditions}
\end{table}

{The target channel parameters are generated differently for \ac{LoS} and \ac{NLoS} conditions. For \ac{LoS} target channel, all parameters (\ac{AA} at the \ac{Rx}, \ac{AD} at the \ac{Tx}, \ac{AA} and \ac{AD} at the target, and the propagation delay) are derived deterministically from the \ac{Tx}-target-\ac{Rx} geometry. In contrast, for \ac{NLoS} conditions, we employ a hybrid clustering approach that combines the stochastic and deterministic clusters. Our framework uses deterministic clusters, representing Type I \acp{EO} to ensure strict spatiotemporal consistency. These clusters are modeled through geometric computation of all propagation paths (\ac{Tx}-target, target-\ac{Rx}, and target-cluster interactions) based on the physical environment's 3D configuration. On the other hand, stochastic clusters serve as an optional addition in \ac{NLoS} target channel to represent clutter/interference that degrades the sensing performance. The \ac{NLoS} cluster generation process begins with standard \ac{TR}38.901 \cite{tr} procedures for both links. Subsequently, among the total generated clusters, a subset becomes deterministic clusters with enforced geometric consistency and tailored scattering properties, while the remainder remains stochastic.}

The \ac{NLoS} target channel implementation incorporates absolute propagation delays for both stochastic and deterministic clusters to ensure a physically accurate superposition of all \ac{CIR} components at the \ac{Rx}. This requirement arises from the need to preserve correct temporal relationships between paths originating from targets, target channel clusters, and the background channel. The methodology, however, differs between cluster types. For stochastic clusters, absolute delays are generated using the procedural framework defined in \ac{TR}38.901 \cite{tr}, Section 7.6.9. In contrast, deterministic clusters possess inherently absolute propagation delays because of geometric consistency, where each path's delay is explicitly calculated from the positions of the \ac{Tx}, \ac{Rx}, and the cluster within a unified coordinate system.
\subsection{Background Channel Characterization and Modeling}\label{sec2b}
The background channel (as given in Fig. \ref{prop_cond}) consists of a \ac{LoS} component, and stochastic \ac{NLoS} components modeled via cluster-based approach in \ac{TR}38.901\cite{tr}. As advocated by current \ac{ISAC} frameworks \cite{3gpp_R1-2504945}, this methodology provides sufficient characterization of environmental scattering components without requiring precise geometric positioning of clusters cite{7248689}. This permits the adoption of the stochastic \ac{NLoS} \ac{CM} framework specified in \ac{TR}38.901 \cite{tr} where clusters are represented through statistical parameters, whereby each multipath component is characterized by statistical parameters, e.g., power, delay, Doppler shift, \ac{AA} \& \ac{AD}. Also, the background channel incorporates absolute delays for \ac{NLoS} clusters. 
\vspace{-0.1cm}
\section{Bistatic ISAC \acl{CM}}\label{sec3}
{The proposed \ac{ISAC} channel model framework is designed to incorporate the four bistatic sensing modes specified by \ac{3GPP}, that are, \ac{TRP}-\ac{TRP}, \ac{TRP}-\ac{UE}, \ac{UE}-\ac{TRP}, and \ac{UE}-\ac{UE} \cite{3gpp_R1-2504945}. Note that a \ac{TRP} is a \ac{BS}. At the most basic level, these four bistatic \ac{ISAC} sensing modes primarily differ based on the mobility and roles of the \ac{Tx} and \ac{Rx}. For instance, the \ac{TRP}-\ac{TRP} mode involves two fixed \acp{BS} for network-controlled sensing, while the \ac{UE}-\ac{UE} mode enables decentralized sensing between mobile devices. A comprehensive framework must therefore be adaptable to these distinct scenarios.}

\begin{figure*}[tb]
  \centering
  \includegraphics[trim={1mm 1mm 1.2mm 1mm},clip,scale=0.85]{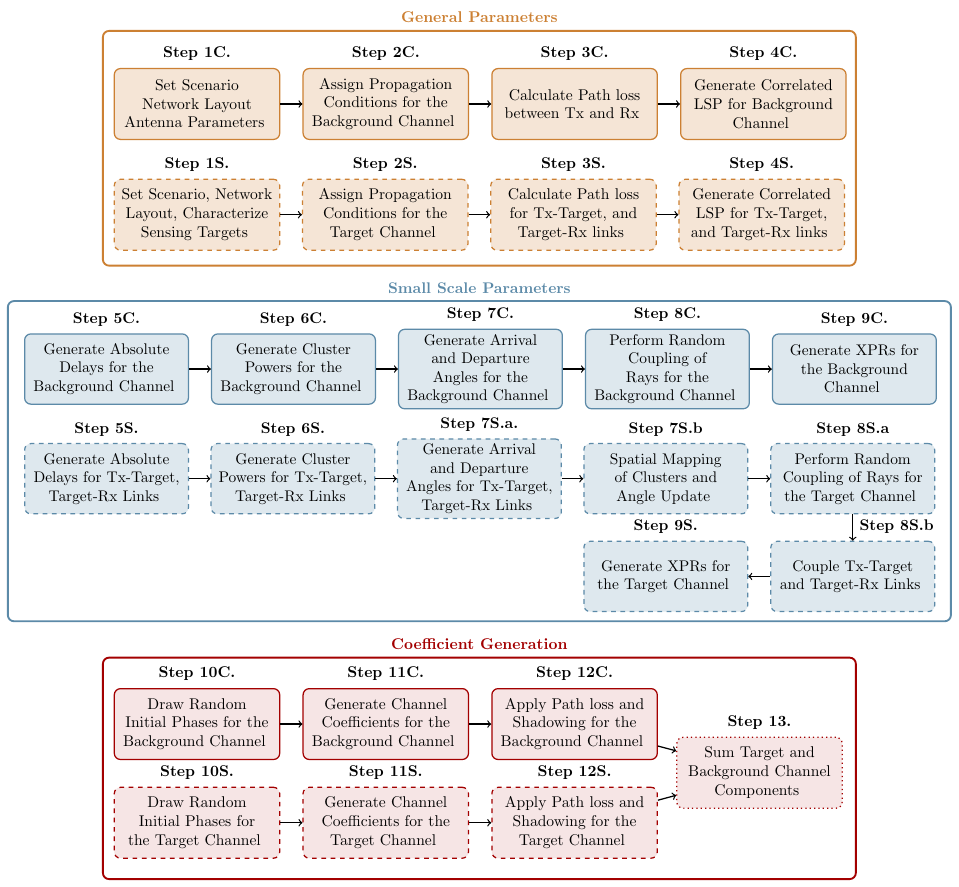}
  \caption{Proposed bistatic \ac{GBSM} \ac{ISAC} \ac{CM} framework. The processing steps for the background channel are assigned a "C" suffix and are represented by solid-outline rectangles, whereas steps for the target channel use an "S" suffix and are delineated with dashed outlines.}
  \label{framework}
\end{figure*}

{As depicted in Fig. \ref{framework}, the proposed model follows a structured approach similar to \ac{TR}38.901 \cite{tr}, partitioning the modeling process into three core components, that are, (1) the general parameters, (2) small-scale parameters, and (3) coefficient generation. The following sections provide a detailed explanation of each component within our unified \ac{ISAC} \ac{CM} framework.}
\subsection{General Parameters}\label{sec3a}
{The general parameters define the propagation environment, network configuration, and array characteristics to ensure consistent modeling for both the background and target channels.  It begins with selecting a standard scenario (e.g., \ac{UMa}, \ac{UMi}, or \ac{InF}), defining the global and spherical coordinate systems (\(\theta\), \(\phi\)), and specifying scenario-specific sensing requirements. Following this, the network topology and target characteristics are defined, which include: (1) \ac{Tx}/\ac{Rx} positions and mobility based on sensing modes, (2) the number of targets, \(L\), (3) target \ac{RCS}, \(\sigma\) and the number of \acp{SP}, \(K\), (4) the position vector for \(k\)th \ac{SP} of \(l\)th target, \(\boldsymbol{d}_{l,k}\), where \(l = 1,2,\cdots, L\) and \(k = 1,2,\cdots,K\), (5) the target velocity (assuming that all the \acp{SP} have the same velocity), (6) number of deterministic clusters in \ac{Tx}-target and target-\ac{Rx} links, (7) the \ac{RCS} of the deterministic clusters, and (8) the motion characteristics of the deterministic clusters. The positions of target cluster(s) and deterministic clusters if known or generated randomly, should be explicitly defined. The system center frequency \(f_c\) completes the parameterization.}

{\ac{LoS} angles are computed for both the background channel and the target channel considering the \ac{Tx}, target, and \ac{Rx} positions. For the background channel, \ac{LoS} \ac{AoA} and \ac{ZoA} at the \ac{Rx}, i.e., \(\phi_\mathrm{LoS,AoA}\), \(\theta_\mathrm{LoS,ZoA}\), \ac{AoD} and \ac{ZoD} at the \ac{Tx}, i.e., \(\phi_\mathrm{LoS,AoD}\), \(\theta_\mathrm{LoS,ZoD}\). For the target channel, angles along the \ac{Tx}-target and target-\ac{Rx} paths are calculated, which include \ac{AoA} and \ac{ZoA} at the Rx and \ac{AoD} and \ac{ZoD} at the Tx along the Tx-target and target-Rx links in the target channel, i.e., \(\phi_{l,k,\mathrm{AoA}}\), \(\theta_{l,k,\mathrm{ZoA}}\), \(\phi_{l,k,\mathrm{AoD}}\), and \(\theta_{l,k,\mathrm{ZoD}}\). This parametrization completes Step 1C and Step 1S.}

{The background and the target  channel propagation conditions (\ac{LoS}/\ac{NLoS}) are assigned independently using \ac{3GPP} probability models defined in different \acp{TR}. The background channel condition (Step 2C) is determined for the \ac{Tx}-\ac{Rx} link, while the target channel conditions (Step 2S) are independently evaluated for both \ac{Tx}-target and target-\ac{Rx} links \cite{3gpp_R1-2504945}. Afterwards, the path loss is obtained for the background channel (Step 3C) and both links in the target channel (Step 3S) using standardized models in \acp{TR}. Finally, correlated large-scale parameters, that are, \ac{DS}, \ac{AS}, \ac{SF}, \(K\)-factor are generated for the background channel (Step 4C) and independently for both \ac{Tx}-target and target-\ac{Rx} links (Step 4S).}
\subsection{Small Scale Parameters}\label{sec3b}
\textbf{Absolute Delay Generation and Cluster Classification}: The generation of small-scale parameters initiates with Steps 5C and 5S for the background and target channels, respectively, following the \ac{TR}38.901 \cite{tr} procedure to produce relative cluster delays. For the background \ac{Tx}-\ac{Rx} link, this yields \(N\) clusters with delays \(\tau'_n\) (\(n = 0,1,\cdots,N-1\)), where the first cluster represents the \ac{LoS} path and is set to zero. For the target channel, \(N_1\) and \(N_2\) clusters are independently generated for the \ac{Tx}-target and target-\ac{Rx} links, with delays \(\tau'_{n_1}\) (\(n_1 = 0,1,\cdots,N_1-1\)) and \(\tau'_{n_2}\) (\(n_2 = 0,1,\cdots,N_2-1\)), respectively, also with their first clusters set to zero. 

Here, \(N'_1 = N_1-1\) and \(N'_2 = N_2-1\). Subsequently, \ac{NLoS} clusters are categorized as either deterministic or stochastic. The \ac{Tx}-target link clusters are divided into \(D_1\) deterministic and \(S_1\) stochastic clusters, while the target-\ac{Rx} link clusters comprise \(D_2\) deterministic and \(S_2\) stochastic clusters, where \(S_1 = N_1' - D_1\) and \(S_2 = N_2' - D_2\). After extracting the deterministic clusters (indexed \(p_1 = 0,1,\cdots,D_1-1\) and \(p_2 = 0,1,...,D_2-1\)), the remaining stochastic clusters are reindexed to \(n_1 = 0,1,\cdots,S_1-1\) and \(n_2 = 0,1,\cdots,S_2-1\). This classification is critical as the absolute delay modeling differs between the deterministic and stochastic clusters. It is important to note that the conventional \ac{TR}38.901 \cite{tr} framework defines a methodology for absolute delays but operates primarily on a relative delay basis.

The calculation of absolute propagation delays is critical for \ac{ISAC} \ac{CM}. For the background channel, the baseline is established by the \ac{LoS} delay between the \ac{Tx} and \ac{Rx}, \(\tau_\mathrm{LoS} = d_{\mathrm{Tx,Rx}} / c\). The absolute delay for the \ac{LoS} cluster (\(n=0\)) is equal to the baseline delay, \(\tau_0 = \tau_\mathrm{LoS}\). For \ac{NLoS} clusters (\(n > 0\)), an excess delay term \(\Delta \tau\) is incorporated, resulting in a total absolute delay of \(\tau_n = \tau'_n + \tau_\mathrm{LoS} + \Delta \tau\). The final ray delays within each cluster, denoted \(\tau_{n, m}\) where \(m = 1,2,\cdots,M\) is the ray index, are generated by applying the intra-cluster delay distribution procedure specified in Section 7.6.2.2 of \ac{TR}38.901 \cite{tr} to these absolute cluster delays \(\tau_n\). Here, \(M\) is the number of rays in a cluster.

In contrast, the target channel requires a separate computation for the \ac{Tx}-target and target-\ac{Rx} segments. Each segment has its own LoS baseline, i.e., \(\tau^{(1)}_\mathrm{LoS} = d_{\mathrm{Tx,target}} / c\) for the \ac{Tx}-target link and \(\tau^{(2)}_\mathrm{LoS} = d_{\mathrm{target,Rx}} / c\) for the target-\ac{Rx} link. These baselines are then combined with cluster-specific delays that differ based on cluster type. Stochastic clusters in the target channel incorporate an excess delay (\(\Delta \tau^{(1)}, \Delta \tau^{(2)}\)), yielding delays \(\tau_{n_1} = \tau'_{n_1} + \tau^{(1)}_\mathrm{LoS} + \Delta \tau^{(1)}\) for the \ac{Tx}-target link and \(\tau_{n_2} = \tau'_{n_2} + \tau^{(2)}_\mathrm{LoS} + \Delta \tau^{(2)}\). Deterministic clusters, governed by strict geometric consistency, do not require the stochastic excess delay, resulting in geometrically derived delays \(\tau_{p_1} = \tau'_{p_1} + \tau^{(1)}_\mathrm{LoS}\) for the \ac{Tx}-target link and \(\tau_{p_2} = \tau'_{p_2} + \tau^{(2)}_\mathrm{LoS}\) for the target-\ac{Rx} link. Finally, the intra-cluster ray generation procedure is applied, producing the final sets of path delays for the target channel's stochastic (\(\tau_{n_1,m_1}\), \(\tau_{n_2,m_2}\)) and deterministic (\(\tau_{p_1,k_1}\), \(\tau_{p_2,k_2}\)) clusters.

\textbf{Cluster Power Generation}: {Subsequently, cluster powers are generated following the \ac{TR}38.901 \cite{tr} procedure in Steps 6C and 6S. For the background channel, powers \(P_n\) are derived based on the \ac{DS}, \(\tau'_n\), \ac{SF}, and \(K^\mathrm{back}\). For the target channel, powers are computed independently for each link, cluster powers \(P_{n_1}\) for the \ac{Tx}-target link use parameters \(\tau'_{n_1}\), \(K^\mathrm{tar}_{1}\), and corresponding \ac{DS} and \ac{SF} values, while powers \(P_{n_2}\) for the target-\ac{Rx} link use \(\tau'_{n_2}\), \(K^\mathrm{tar}_{2}\), and its associated parameters. Power management employs distinct thresholding strategies; the standard \(-25\) dB threshold from \ac{TR}38.901 \cite{tr} is applied for initial cluster removal in all links, while a relaxed \(-40\) dB threshold as agreed by \ac{3GPP} \cite{3gpp_R1-2504945} is used during post-concatenation path dropping to accommodate the cascaded nature of bistatic sensing. For ray-level powers, \ac{NLoS} clusters in the background channel feature equal power distribution among \(M\) rays, yielding \(P_{n,m}\). The target channel requires distinct treatments, i.e.,  stochastic clusters generate ray powers \(P_{n_1,m_1}\) and \(P_{n_2,m_2}\) for the \ac{Tx}-target and target-\ac{Rx} links, respectively, while deterministic clusters produce \ac{SP} powers \(P_{p_1,k_1}\) and \(P_{p_2,k_2}\) following a similar procedure as for stochastic clusters.}

\textbf{Cluster Angle Generation}: The \ac{AD} at the \ac{Tx} and \ac{AA} at the \ac{Rx} for each ray within the \ac{NLoS} clusters are generated in Step 7C. For the \(m\)th ray in the \(n\)th cluster, the \ac{AD} at the \ac{Tx} are characterized by the \ac{AoD}, \(\phi_{n,m,\mathrm{AoD}}\) and \ac{ZoD}, \(\theta_{n,m,\mathrm{ZoD}}\), while the \ac{AA} at \ac{Rx} are characterized by the \ac{AoA}, \(\phi_{n,m,\mathrm{AoA}}\) and \ac{ZoA}, \(\theta_{n,m,\mathrm{ZoA}}\). This process follows the \ac{TR}38.901 \cite{tr} methodology, where the \ac{AD} are statistically generated based on the \ac{ASD} and \ac{ZSD}, and the \ac{AA} are generated based on the \ac{ASA} and \ac{ZSA} obtained for the \ac{Tx}-\ac{Rx} link within the background channel.

{For the target channel, angles are obtained for both the \ac{Tx}-target and target-\ac{Rx} links. In the \ac{Tx}-target link, the azimuth and zenith \ac{AD} (\(\phi_{n_1,m_1,\mathrm{AoD}}\), \(\theta_{n_1,m_1,\mathrm{ZoD}}\)) are generated at the \ac{Tx}, with corresponding \ac{AA} (\(\phi_{n_1,m_1,\mathrm{AoA}}\), \(\theta_{n_1,m_1,\mathrm{ZoA}}\)) at the target. Similarly, for the target-\ac{Rx} link, \ac{AD} (\(\phi_{n_2,m_2,\mathrm{AoD}}\), \(\theta_{n_2,m_2,\mathrm{ZoD}}\)) are generated at the target, with \ac{AA} (\(\phi_{n_2,m_2,\mathrm{AoA}}\), \(\theta_{n_2,m_2,\mathrm{ZoA}}\)) obtained at the \ac{Rx}. For deterministic clusters, the corresponding angles are denoted as \(\phi_{p_1,k_1,\mathrm{AoD}}\), \(\theta_{p_1,k_1,\mathrm{ZoD}}\), \(\phi_{p_1,k_1,\mathrm{AoA}}\), \(\theta_{p_1,k_1,\mathrm{ZoA}}\) in the \ac{Tx}-target link, and \(\phi_{p_2,k_2,\mathrm{AoD}}\), \(\theta_{p_2,k_2,\mathrm{ZoD}}\), \(\phi_{p_2,k_2,\mathrm{AoA}}\), \(\theta_{p_2,k_2,\mathrm{ZoA}}\) in the target-\ac{Rx} link.}

\textbf{Spatial Mapping of Deterministic Clusters}: {At this stage, only the classification between stochastic and deterministic clusters is complete, while the spatial mapping for deterministic clusters remains unprocessed. For the background channel, Step 8C performs random coupling of rays for both azimuth and zenith angles as defined in \ac{TR}38.901 \cite{tr}. For the target channel, Step 7Sa involves spatial mapping of deterministic clusters, where each cluster (or individual rays within it) are mapped to specific spatiotemporal locations using either \ac{AA} or \ac{AD} combined with cluster delay via the law of cosines. When cluster position is determined from one angle set (e.g., \ac{AD}), the complementary set (\ac{AA}) must be adjusted to ensure spatiotemporal consistency, realigning them with the determined 3D coordinates, which follows an approach similar to QuadRiGa.}

\begin{figure}[tb]
    \centering
    \includegraphics[trim={5mm 1mm 1mm 1mm},clip,width=0.45\textwidth]{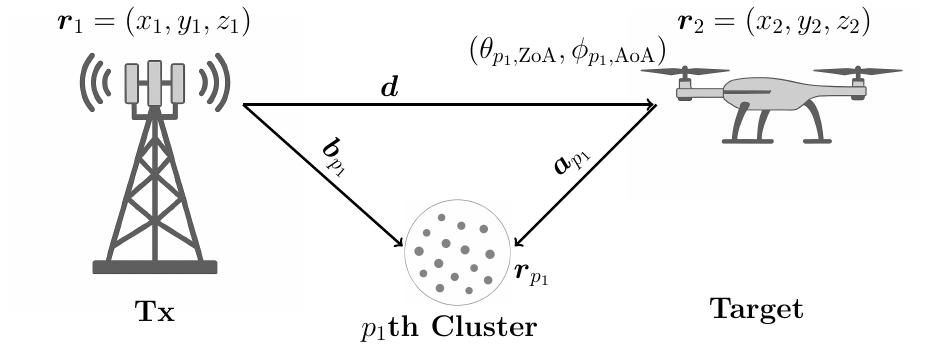} 
    \caption{Spatial mapping of deterministic clusters.}
    \label{spatia_mapping} 
\end{figure}

Consider the scenario illustrated in Fig. \ref{spatia_mapping}, depicting the 
\(p_1\)th cluster in the \ac{Tx}-target link. The \ac{Tx} and target positions are defined by Cartesian coordinates, \(\boldsymbol{r}_1\) and \(\boldsymbol{r}_2\), respectively, with their separation given by \(\boldsymbol{d} = \boldsymbol{r}_2 - \boldsymbol{r}_1\)  and distance \(d = \Vert \boldsymbol{d}\Vert\). The vector from the \ac{Tx} to the \(p_1\)th cluster is \(\boldsymbol{b}_{p_1}=\hat{\boldsymbol{b}}_{p_1}\vert \boldsymbol{b}_{p_1}\vert\), where \(\hat{\boldsymbol{b}}_{p_1}\) is the spherical unit vector and \(\vert \boldsymbol{b}_{p_1}\vert\) is its magnitude. Similarly, the vector from the target to the \(p_1\)th cluster is \(\boldsymbol{a}_{p_1}=\hat{\boldsymbol{a}}_{p_1}\vert \boldsymbol{a}_{p_1}\vert\), where \(\hat{\boldsymbol{a}}_{p_1}\) is the corresponding spherical unit vector and \(\vert \boldsymbol{a}_{p_1}\vert\) is its length. The azimuth and zenith {\ac{AA}} at the target for this \(p_1\)th cluster are \(\phi_{p_1,\mathrm{AoA}}\) and \(\theta_{p_1,\mathrm{ZoA}}\), respectively, and its absolute delay is \(\tau_{p_1}\). The cluster's distance is derived from the delay as \(d_{p_1} = \tau_{p_1} c\). The spherical unit vector 
\(\hat{\boldsymbol{a}}_{p_1}\) is determined from the arrival angles as \(\hat{\boldsymbol{a}}_{p_1} =\boldsymbol{s}\left(\theta_{p_1,\mathrm{ZoA}},\phi_{p_1,\mathrm{AoA}}\right)\). Applying the law of cosines, the magnitude \(\vert \boldsymbol{a}_{p_1}\vert\) {is computed as \(
|\boldsymbol{a}_{p_1}| = \frac{d_{p_1}^2 - d^2}{2(d_{p_1} - \boldsymbol{d}^\mathrm{T} \hat{\boldsymbol{a}}_{p_1})}\)}.

Finally, the cluster's spatial coordinates are obtained as \(\boldsymbol{r}_{p_1} = \boldsymbol{r}_2 + \hat{\boldsymbol{a}}_{p_1} \vert \hat{\boldsymbol{a}}_{p_1}\vert \). With the 3D coordinates of the \(p_1\) th cluster now determined, the {\ac{AD}} at the \ac{Tx} are recalculated to ensure geometric consistency with the cluster's exact spatial position. This adjustment aligns the azimuth (\(\phi_{p_1,\mathrm{AoD}}\)) and zenith (\(\theta_{p_1,\mathrm{ZoD}}\)) {\ac{AD}} with the derived cluster location. A similar approach can be adopted for the target-\ac{Rx} link having clusters indexed with \(p_2\). Moreover, same method can be used to obtain the target position vector if generated through stochastic target placement.

\textbf{Ray Coupling and \ac{XPR} Generation}: {Step 8Sa performs random ray coupling for stochastic clusters in both the \ac{Tx}-target and target-\ac{Rx} links, following the \ac{TR}38.901 \cite{tr} methodology. Step 8Sb extends this coupling to include both stochastic and deterministic clusters from both \ac{Tx}-target and target-\ac{Rx} links, systematically traversing all possible multipath combinations at the target: (1) \ac{LoS}-\ac{LoS} coupling, (2) \ac{LoS}-\ac{NLoS} coupling, and (3) \ac{NLoS}-\ac{NLoS} coupling. In Step 9C, \acp{XPR}, ($\kappa_{n,m}$) are generated per \ac{TR}38.901 \cite{tr}'s statistical procedure. For Step 9S, \ac{XPR} modeling covers both statistical and deterministic clusters across all Table \ref{propation_conditions} propagation cases:  (1) $\kappa_{n_1,m_1,1}$ (\ac{LoS} target-\ac{Rx}/\ac{NLoS} \ac{Tx}-target), (2) $\kappa_{1,n_2,m_2}$ (\ac{LoS} \ac{Tx}-target/\ac{NLoS} target-\ac{Rx}), and (3) $\kappa_{n_1,m_1,n_2,m_2}$ (\ac{NLoS} in both links) for statistical clusters, with analogous $\kappa_{p_1,k_1,1}$, $\kappa_{1,p_2,k_2}$, and $\kappa_{p_1,k_1,p_2,k_2}$ for deterministic cases. Statistical \acp{XPR} follow standard stochastic modeling, while deterministic \acp{XPR} incorporate \ac{EO} geometry and material properties.}
\subsection{Coefficient Generation}
{The initial phase generation follows \ac{TR}38.901 \cite{tr} specifications for both background (Step 10C) and both links in the target channels (Step 10S). For the background channel, each ray $m$ within cluster $n$ is assigned four random initial phases $\boldsymbol{\Phi}_{n,m}^{\mathrm{pol}}$ (where $\mathrm{pol} \in \{\theta\theta, \theta\phi, \phi\theta, \phi\phi\}$) drawn uniformly from $(-\pi,\pi]$. The target channel extends this to all propagation conditions with statistical clusters \cite{587631} require three phase sets, that are, $\Phi_{n_1,m_1,1}^{\mathrm{pol}}$ for \ac{NLoS} \ac{Tx}-target/\ac{LoS} target-\ac{Rx}, $\Phi_{1,n_2,m_2}^{\mathrm{pol}}$ for \ac{LoS} \ac{Tx}-target/\ac{NLoS} target-\ac{Rx}, and $\Phi_{n_1,m_1,n_2,m_2}^{\mathrm{pol}}$ for \ac{NLoS} \ac{Tx}-target/\ac{NLoS} target-\ac{Rx}. The deterministic clusters follow an identical structure with $(p_i,k_i)$ replacing $(n_i,m_i)$ indices, generating $\Phi_{p_1,k_1,1}^{\mathrm{pol}}$, $\Phi_{1,p_2,k_2}^{\mathrm{pol}}$, and $\Phi_{p_1,k_1,p_2,k_2}^{\mathrm{pol}}$ for the respective propagation cases.}

The \ac{ISAC} \ac{CM} procedure progresses to coefficient generation in Step 11C. and Step 11S. In the sequel, we first derive the complex \ac{CC} for the background channel, and then for the target channel. 
\subsubsection{Background Channel}
The background channel models the aggregate of all propagation paths between the \ac{Tx} and \ac{Rx} that do not interact with the sensing target. It comprises (i) the \ac{LoS} path and (ii) stochastic \ac{NLoS} components. A generalized expression for generating \ac{CC} is given by:
\begin{equation}\label{cc_background}
H_{u,s,\boldsymbol{\mathcal{Q}}}^{\zeta}(t) = \Gamma^{\zeta} \mathbf{F}_{\mathrm{rx},u}^\mathrm{T}\left(\boldsymbol{\Omega}^{\zeta}_{\mathrm{rx}}\right) \boldsymbol{\mathcal{C}}^{\zeta} \mathbf{F}_{\mathrm{tx},s}\left(\boldsymbol{\Omega}^{\zeta}_{\mathrm{tx}}\right)\!\! \prod_{i \in \{\mathrm{tx}, \mathrm{rx}\}}\!\! \mathcal{L}_i \, \mathcal{D}_i(t),
\end{equation}
where \(\zeta = \mathrm{LoS, back}\) for \ac{LoS} or \(\zeta = \mathrm{NLoS, back}\) for \ac{NLoS} propagation condition for the background channel. \(\boldsymbol{\mathcal{Q}}\) defines the  propagation path index, e.g. \(\boldsymbol{\mathcal{Q}} = \{1\}\) for \ac{LoS} and \(\boldsymbol{\mathcal{Q}} = \{n,m\}\) for the \ac{NLoS} condition. The model parameters are defined as follows: \(\Gamma^\zeta\) can be the complex channel gain, phase term or combination of both. \(\mathbf{F}_{\mathrm{tx},s}\left(\boldsymbol{\Omega}_{\mathrm{tx}}^{\zeta}\right)\) and \(\mathbf{F}_{\mathrm{rx},u}\left(\boldsymbol{\Omega}_{\mathrm{rx}}^{\zeta}\right)\) are the field patterns of the \(s\)th transmit and \(u\)th receive antenna elements, respectively, evaluated at the \ac{AD}, \(\boldsymbol{\Omega}_{\mathrm{tx}}^{\zeta}\) and \ac{AA}, \(\boldsymbol{\Omega}_{\mathrm{rx}}^{\zeta}\). The polarimetric properties are characterized by the \ac{CPM}, \(\boldsymbol{\mathcal{C}}^{\zeta}\). The spatial location phase term, \(\mathcal{L}_j\) accounts for the phase shift induced by the geometry of the antenna array elements, while the Doppler term, \(\mathcal{D}_i(t)\) models the phase shift resulting from the relative motion of \ac{Tx} and/or the \ac{Rx}. The subsequent sections provide explicit definitions for these parameters for \ac{LoS} and \ac{NLoS} components. We define the spherical unit direction vector as \(\boldsymbol{s}(\theta,\phi) = [\sin\theta \cos\phi,\ \sin\theta \sin\phi,\ \cos\theta]^\mathrm{T}\), where \(\theta\) and \(\phi\) are the zenith and azimuth angles, respectively. The velocity vector is \(\boldsymbol{v}(v,\theta,\phi) = v \boldsymbol{s}(\theta,\phi)\), with \(v\) denoting velocity magnitude.
\paragraph{\ac{LoS} Component}
The \ac{LoS} component represents the direct path between the \ac{Tx} and \ac{Rx} and is modeled in accordance with \ac{TR}38.901 \cite{tr}. Its \ac{CC} are derived using eq. \eqref{cc_background} by setting \(\boldsymbol{\mathcal{Q}} =\{1\}\) (convention used in \ac{TR}38.901 \cite{tr}) and \(\zeta = \mathrm{LoS,back}\). For the \ac{LoS} background channel, \(\Gamma^{\mathrm{LoS,back}}\) represents the phase term which is given as \(\Gamma^{\mathrm{LoS,back}} = \exp\left(-j2\pi d_\mathrm{Tx,Rx}/\lambda_0\right)\), where
\(\lambda_0\) is the carrier wavelength. The antenna field patterns, \(\mathbf{F}_{\mathrm{tx},s}\) and \(\mathbf{F}_{\mathrm{rx},u}\), are evaluated at \(\boldsymbol{\Omega}_\mathrm{tx}^{\mathrm{LoS,back}}=\{\theta_\mathrm{LoS,ZoD},\phi_\mathrm{LoS,AoD}\}\) and \(\boldsymbol{\Omega}_\mathrm{rx}^{\mathrm{LoS,back}}=\{\theta_\mathrm{LoS,ZoA},\phi_\mathrm{LoS,AoA}\}\). The \ac{CPM} is given as \(\boldsymbol{\mathcal{C}}^\mathrm{LoS,back} = \begin{bmatrix}1 & 0\\ 0 & -1\end{bmatrix}\). The spatial location phase term for a given antenna element position \(\boldsymbol{d}\) and unit direction vector \(\hat{\boldsymbol{r}}\) is defined as \(\mathcal{L}(\hat{\boldsymbol{r}}, \boldsymbol{d}) = \exp\left(j2\pi(\hat{\boldsymbol{r}}^\mathrm{T} \cdot \boldsymbol{d})/\lambda_0\right)\). Here, \(\mathcal{L}_\mathrm{tx} = \mathcal{L}(\hat{\boldsymbol{r}}_{\mathrm{tx,LoS}}, \boldsymbol{d}_{\mathrm{tx},s})\) and \(\mathcal{L}_\mathrm{rx} = \mathcal{L}(\hat{\boldsymbol{r}}_{\mathrm{rx,LoS}}, \boldsymbol{d}_{\mathrm{rx},u})\), where \(\hat{\boldsymbol{r}}_{\mathrm{tx,LoS}}= \boldsymbol{s}(\theta_\mathrm{LoS,ZoD},\phi_\mathrm{LoS,AoD})\) and \(\hat{\boldsymbol{r}}_{\mathrm{rx,LoS}} = \boldsymbol{s}(\theta_\mathrm{LoS,ZoA},\phi_\mathrm{LoS,AoA})\). The Doppler term for a given velocity vector, \(\boldsymbol{v}\) is defined as \(\mathcal{D}(\hat{\boldsymbol{r}}, \boldsymbol{v}, t) = \exp\left(j2\pi(\hat{\boldsymbol{r}}^\mathrm{T} \cdot \boldsymbol{v})t/\lambda_0\right)\). For \ac{LoS} component, we have \(\mathcal{D}_\mathrm{tx}(t) = \mathcal{D}(\hat{\boldsymbol{r}}_{\mathrm{tx,LoS}}, \boldsymbol{v}_\mathrm{tx}, t)\) and \(\mathcal{D}_\mathrm{rx}(t) = \mathcal{D}(\hat{\boldsymbol{r}}_{\mathrm{rx,LoS}}, \boldsymbol{v}_\mathrm{rx}, t)\). The velocity vectors are \(\boldsymbol{v}_\mathrm{tx} = \boldsymbol{v}(v_\mathrm{tx},\theta_\mathrm{tx},\phi_\mathrm{tx})\) for the \ac{Tx} and \(\boldsymbol{v}_\mathrm{rx} = \boldsymbol{v}(v_\mathrm{rx},\theta_\mathrm{rx},\phi_\mathrm{rx})\) for the \ac{Rx}, where \(v_\mathrm{tx}\) and \(v_\mathrm{rx}\) are the velocity magnitudes, while \(\theta_\mathrm{tx}, \phi_\mathrm{tx}\) and \(\theta_\mathrm{rx}, \phi_\mathrm{rx}\) are the zenith and azimuth angles of the travel for \ac{Tx} and \ac{Rx}, respectively. Note that the classical \ac{TR}38.901 \cite{tr} does not incorporate the velocity of the \ac{Tx} as it is assumed to be static. However, for \ac{ISAC} \ac{CM} it is mandatory to consider the velocity for the both the \ac{Tx} and \ac{Rx} as both can be either the \ac{BS} or \ac{UE}. 
\paragraph{\ac{NLoS} Component}
The \ac{NLoS} component aggregates contributions from multiple stochastic clusters, each comprising several rays, as defined by \ac{TR}38.901 \cite{tr}. The \ac{CC} for the \(m\)th ray within the \(n\)th cluster is obtained from the generalized structure in eq. \eqref{cc_background} with \(\boldsymbol{\mathcal{Q}} = \{n,m\}\) and \(\zeta = \mathrm{NLoS,back}\). In this case, \(\Gamma^{\mathrm{NLoS,back}}\) is the ray power, \(\sqrt{P_{n,m}}\). The antenna field patterns are evaluated at ray-specific angles, that are, \(\boldsymbol{\Omega}_\mathrm{tx}^{\mathrm{NLoS,back}}=\{\theta_{n,m,\mathrm{ZoD}},\phi_{n,m,\mathrm{AoD}}\}\) and \(\boldsymbol{\Omega}_\mathrm{rx}^{\mathrm{NLoS,back}}=\{\theta_{n,m,\mathrm{ZoA}},\phi_{n,m,\mathrm{AoA}}\}\). The \ac{CPM} is given as \(\boldsymbol{\mathcal{C}}^{\mathrm{NLoS,back}} = \boldsymbol{\mathcal{C}}(\kappa_{n,m},\boldsymbol{\Phi}_{n,m}^\mathrm{pol})=
\begin{bmatrix} 
\exp(j\Phi^{\theta\theta}_{n,m}) & \sqrt{\kappa_{n,m}^{-1}}\exp(j\Phi^{\theta\phi}_{n,m}) \\ 
\sqrt{\kappa_{n,m}^{-1}}\exp(j\Phi^{\phi\theta}_{n,m}) & \exp(j\Phi^{\phi\phi}_{n,m}) 
\end{bmatrix}\), where \(\kappa_{n,m}\) is the \ac{XPR} and \(\boldsymbol{\Phi}_{n,m}\) contains random initial phases. The spatial location term for \ac{NLoS} case uses the same functional form but with ray-dependent unit direction vectors. Thus, \(\mathcal{L}_\mathrm{tx} = \mathcal{L}(\hat{\boldsymbol{r}}_{\mathrm{tx},n,m}, \boldsymbol{d}_{\mathrm{tx},s})\) and \(\mathcal{L}_\mathrm{rx} = \mathcal{L}(\hat{\boldsymbol{r}}_{\mathrm{rx},n,m}, \boldsymbol{d}_{\mathrm{rx},u})\), where \(\hat{\boldsymbol{r}}_{\mathrm{tx},n,m} = \boldsymbol{s}(\theta_{n,m,\mathrm{ZoD}},\phi_{n,m,\mathrm{AoD}})\) and \(\hat{\boldsymbol{r}}_{\mathrm{rx},n,m} = \boldsymbol{s}(\theta_{n,m,\mathrm{ZoA}},\phi_{n,m,\mathrm{AoA}})\). Similarly, the Doppler terms for the \ac{Tx} and the \ac{Rx} are given as \(\mathcal{D}_\mathrm{tx}(t) = \mathcal{D}(\hat{\boldsymbol{r}}_{\mathrm{tx},n,m}, \boldsymbol{v}_\mathrm{tx}, t)\) and \(\mathcal{D}_\mathrm{rx}(t) = \mathcal{D}(\hat{\boldsymbol{r}}_{\mathrm{rx},n,m}, \boldsymbol{v}_\mathrm{rx}, t)\). 

The background channel \ac{CIR} is given as:
\begin{equation}\label{LOS_cir_background}
\boldsymbol{\mathcal{H}}_{u,s}^\mathrm{back}(t,\tau) =\gamma\mathbf{H}_{u,s}^\mathrm{LoS,back}(t,\tau) + \tilde{\gamma} \mathbf{H}_{u,s}^\mathrm{NLoS,back}(t,\tau)
\end{equation}
where \(\gamma =\sqrt{{K^\mathrm{back}}/{K^\mathrm{back}+1}}\) and \(\tilde{\gamma} = \sqrt{{1}/{K^\mathrm{back}+1}}\) are the Rician \(K\)-factor weights. The \ac{LoS} component is given as \(\mathbf{H}_{u,s}^\mathrm{LoS,back}(t,\tau) = \left[0,\cdots,H_{u,s}^\mathrm{LoS,back}(t)\delta(\tau-\tau_{u,s}),\cdots,0\right]\) and the \ac{NLoS} component as \(\mathbf{H}_{u,s}^\mathrm{NLoS,back}(t,\tau) = \sum_{n=1}^{N} \sum_{m=1}^{M} \left[H_{u,s,n,m}^\mathrm{NLoS,back}(t)\delta(\tau-\tau_{n,m})\right]\forall~n,m\), where \(\tau_{u,s} = \Vert {\boldsymbol{d}}_{\mathrm{rx},u} - {\boldsymbol{d}}_{\mathrm{tx},s} \Vert / c\) is the \ac{LoS} propagation delay between the \(u\)th receive and \(s\)th transmit antennas, and \(\tau_{n,m}\) are the delays for the \ac{NLoS} components obtained following the procedure in \ac{TR} 38.901 Section 7.6.2.2. Here, each vector represents the \ac{CIR} across discrete delay bins, with non-zero elements only at the specific delay taps corresponding to \(\tau_{u,s}\) and \(\tau_{n,m}\). 
\subsubsection{Target Channel}
The target channel's propagation characteristics are governed by the \ac{Tx}-target and target-\ac{Rx} link states, as categorized in Table \ref{propation_conditions} and illustrated in Fig. \ref{prop_cond}. \ac{LoS} condition is established only when both constituent links exhibit \ac{LoS} propagation; any deviation from this condition in either link results in \ac{NLoS} condition for the target channel. Moreover, the \ac{NLoS} condition may arise from either stochastic or deterministic clusters present in either of the two links.

The \ac{CC} for the target channel between the $u$th receive and $s$th transmit antenna elements are given by the following generalized expression:

\begin{equation}\label{cc_target}
\begin{split}
&H_{u,s,l,k,\boldsymbol{\mathcal{Q}}}^{\zeta}(t) \\
&= \Gamma^{\zeta} \mathbf{F}_{\mathrm{rx},u}^\mathrm{T}\left(\boldsymbol{\Omega}_{\mathrm{rx}}^{\zeta}\right) \boldsymbol{\mathcal{C}}^{\zeta} \mathbf{F}_{\mathrm{tx},s}\left(\boldsymbol{\Omega}_{\mathrm{tx}}^{\zeta}\right) \prod_{i \in \{\mathrm{tx}, \mathrm{rx}\}} \!\!\! \mathcal{L}_i \!\!\! \prod_{j \in \{\mathrm{tx}, \mathrm{rx},\mathrm{clus}\}} \!\!\!\!\!\!\mathcal{D}_j(t),
\end{split}
\end{equation}

For the target channel, the Doppler terms, \(\mathcal{D}_j(t)\) encompass all moving entities (\(j \in \{\mathrm{tx}, \mathrm{rx}, \mathrm{clus}\}\)). The inclusion of the \(\mathcal{D}_\mathrm{clus}\) is critical for modeling the relative motion between the target and the deterministic clusters. Note that the stochastic clusters are modeled as stationary as in \ac{TR}38.901 \cite{tr}.
\paragraph{Case 1: \ac{LoS} Component}

The \ac{LoS} condition manifests when both the \ac{Tx}-target and target-\ac{Rx} links maintain unobstructed propagation paths. This scenario represents a specific instantiation of the generalized model in eq. \eqref{cc_target} with \(\boldsymbol{\mathcal{Q}} = \{1,1\}\) and \(\zeta = \mathrm{LoS,tar}\). Here, \(\Gamma^{\mathrm{LoS,tar}} = \sqrt{\sigma_{l,k}} \alpha(\tilde{d})\) which incorporates the \ac{RCS}, \(\sigma_{l,k}\) of the \(k\)th \ac{SP} of the \(l\)th target and the propagation phase shift, \(\alpha({d}) = \exp(-j2\pi{d}/\lambda_0)\), where \({d} = d_\mathrm{Tx,tar}+d_\mathrm{tar,Rx}\). The angular parameters for the antenna field patterns are \(\boldsymbol{\Omega}_\mathrm{tx}^{\mathrm{LoS,tar}}=\left\{\theta_{l,k,\mathrm{ZoD}},\phi_{l,k,\mathrm{AoD}}\right\}\) and \(\boldsymbol{\Omega}_\mathrm{rx}^{\mathrm{LoS,tar}}=\left\{\theta_{l,k,\mathrm{ZoA}},\phi_{l,k,\mathrm{AoA}}\right\}\). The \ac{CPM} is given as \(\boldsymbol{\mathcal{C}}^{\mathrm{LoS,tar}} = \boldsymbol{\mathcal{C}}^\mathrm{LoS,back}\). The \ac{Tx} spatial phase term is \(\mathcal{L}_\mathrm{tx} = \mathcal{L}(\hat{\boldsymbol{r}}_{\mathrm{tx},l,k},\boldsymbol{d}_{\mathrm{tx},s})\), where the unit directional vector, \(\hat{\boldsymbol{r}}_{\mathrm{tx},l,k} = \boldsymbol{s}(\theta_{l,k,\mathrm{ZoD}},\phi_{l,k,\mathrm{AoD}})\). The \ac{Rx} spatial term is \(\mathcal{L}_\mathrm{rx} = \mathcal{L}(\hat{\boldsymbol{r}}_{\mathrm{rx},l,k},\boldsymbol{d}_{\mathrm{rx},u})\), with \(\hat{\boldsymbol{r}}_{\mathrm{rx},l,k} = \boldsymbol{s}(\theta_{l,k,\mathrm{ZoA}},\phi_{l,k,\mathrm{AoA}})\). The \ac{Tx} Doppler term, \(\mathcal{D}_\mathrm{tx}(t) = \mathcal{D}(\hat{\boldsymbol{r}}_{\mathrm{tx},l,k},\boldsymbol{v}_{\mathrm{tx},l},t)\) employs the relative velocity vector between the \ac{Tx} and the target, \(\boldsymbol{v}_{\mathrm{tx},l} = \boldsymbol{v}_\mathrm{tx}-\boldsymbol{v}_l\). The \ac{Rx} Doppler term, \(\mathcal{D}_\mathrm{rx}(t) = \mathcal{D}(\hat{\boldsymbol{r}}_{\mathrm{rx},l,k},\boldsymbol{v}_{\mathrm{rx},l},t)\) uses the relative velocity, \(\boldsymbol{v}_{\mathrm{rx},l} = \boldsymbol{v}_\mathrm{rx}-\boldsymbol{v}_l\) between the target and \ac{Rx}. For the \ac{LoS} case, the cluster Doppler term is unity, i.e.,  \(\mathcal{D}_\mathrm{clus}(t) = 1\) since no deterministic clusters are present in the propagation path. The target velocity vector, \(\boldsymbol{v}_l = \boldsymbol{v}(v_l,\theta_l,\phi_l)\) is common to all \(K\) \acp{SP} of the \(l\)th target, characterized by speed, \(v_l\) and direction angles, \(\theta_l\), \(\phi_l\).

In addition to the \ac{LoS} components, the target channel also has \ac{NLoS} components. The \ac{NLoS} target components are characterized by the presence of scattering clusters. These scenarios are classified according to whether the clusters are stochastic or deterministic, and whether they occur in the \ac{Tx}-target link, target-\ac{Rx} link, or both. The following sections provide detailed derivations for each distinct \ac{NLoS} condition.
\paragraph{Case 2a: Stochastic \ac{NLoS} Component}

This scenario corresponds to a \ac{LoS} \ac{Tx}-target link with a \ac{NLoS} target-\ac{Rx} link due to stochastic clusters, representing an instantiation of the generalized model in eq. \eqref{cc_target} with \(\boldsymbol{\mathcal{Q}} = \{1,n_2,m_2\}\) and \(\zeta = \mathrm{SNLoS1,tar}\). In this case, \(\Gamma^{\mathrm{SNLoS1,tar}} = \sqrt{\sigma_{l,k}P_{n_2,m_2}}\), where \(P_{n_2,m_2}\) of the \(m_2\)th ray within the \(n_2\)th stochastic cluster in the target-\ac{Rx} link. The antenna field patterns are evaluated at \(\boldsymbol{\Omega}_\mathrm{tx}^{\mathrm{SNLoS1,tar}}=\left\{\theta_{l,k,\mathrm{ZoD}},\phi_{l,k,\mathrm{AoD}}\right\}\) for the \ac{Tx} and \(\boldsymbol{\Omega}_\mathrm{rx}^{\mathrm{SNLoS1,tar}}=\left\{\theta_{n_2,m_2,\mathrm{ZoA}},\phi_{n_2,m_2,\mathrm{AoA}}\right\}\) for the \ac{Rx}. The \ac{CPM} for this case is \(\boldsymbol{\mathcal{C}}^{\mathrm{SNLoS1,tar}} = \boldsymbol{\mathcal{C}}(\kappa_{n_2,m_2},\boldsymbol{\Phi}_{n_2,m_2}^\mathrm{pol})\). The spatial location phase term for the \ac{Tx} is \(\mathcal{L}_\mathrm{tx} = \mathcal{L}(\hat{\boldsymbol{r}}_{\mathrm{tx},l,k},{\boldsymbol{d}}_{\mathrm{tx},s})\) with \(\hat{\boldsymbol{r}}_{\mathrm{tx},l,k}=\boldsymbol{s}(\theta_{l,k,\mathrm{ZoD}},\phi_{l,k,\mathrm{AoD}})\), and for \ac{Rx}, it is \(\mathcal{L}_\mathrm{rx} = \mathcal{L}(\hat{\boldsymbol{r}}_{\mathrm{rx},n_2,m_2},{\boldsymbol{d}}_{\mathrm{rx},u})\), where \(\hat{\boldsymbol{r}}_{\mathrm{rx},n_2,m_2}=\boldsymbol{s}(\theta_{n_2,m_2,\mathrm{ZoA}},\phi_{n_2,m_2,\mathrm{AoA}})\). The \ac{Tx} Doppler term is \(\mathcal{D}_\mathrm{tx}(t) = \mathcal{D}(\hat{\boldsymbol{r}}_{\mathrm{tx},l,k},\boldsymbol{v}_{\mathrm{tx},l},t)\) and the \ac{Rx} Doppler term is \(\mathcal{D}_\mathrm{rx}(t) = \mathcal{D}(\hat{\boldsymbol{r}}_{\mathrm{rx},n_2,m_2},\boldsymbol{v}_\mathrm{rx},t)\). The cluster Doppler term, \(\mathcal{D}_\mathrm{clus}(t) = \mathcal{D}(\hat{\boldsymbol{r}}_{l,n_2,m_2},\boldsymbol{v}_l,t)\) incorporates the target's motion relative to the stochastic cluster, where \(\hat{\boldsymbol{r}}_{l,n_2,m_2}=\boldsymbol{s}(\theta_{n_2,m_2,\mathrm{ZoD}},\phi_{n_2,m_2,\mathrm{AoD}})\).
\paragraph{Case 2b: Deterministic \ac{NLoS} Component}
This scenario features a \ac{LoS} \ac{Tx}-target link with a \ac{NLoS} target-\ac{Rx} link due to deterministic clusters, corresponding to the generalized model in \eqref{cc_target} with \(\boldsymbol{\mathcal{Q}} = \{1,p_2,k_2\}\) and \(\zeta = \mathrm{DNLoS1,tar}\). The \ac{CC} integrate the \ac{LoS} component from the \ac{Tx}-target link with multipath components from deterministic clusters in the target-\ac{Rx} link. For this scenario, \(\Gamma^{\mathrm{DNLoS1,tar}} = \sqrt{\sigma_{l,k}\sigma_{p_2,k_2}P_{p_2,k_2}}\), which incorporates the \ac{RCS} of the deterministic clusters, \(\sigma_{p_2,k_2}\), along with the \ac{SP} power, \(P_{p_2,k_2}\). The antenna field patterns are defined at the angular parameters given as \(\boldsymbol{\Omega}_\mathrm{tx}^{\mathrm{DNLoS1,tar}}=\left\{\theta_{l,k,\mathrm{ZoD}},\phi_{l,k,\mathrm{AoD}}\right\}\) and \(\boldsymbol{\Omega}_\mathrm{rx}^{\mathrm{DNLoS1,tar}} = \left\{\theta_{p_2,k_2,\mathrm{ZoA}},\phi_{p_2,k_2,\mathrm{AoA}}\right\}\). The \ac{CPM} is given as \(\boldsymbol{\mathcal{C}}^{\mathrm{DNLoS1,tar}}=\boldsymbol{\mathcal{C}}\left(\kappa_{p_2,k_2},\boldsymbol{\Phi}_{p_2,k_2}^\mathrm{pol}\right)\). Moreover, \(\mathcal{L}_\mathrm{tx} = \mathcal{L}(\hat{\boldsymbol{r}}_{\mathrm{tx},l,k},{\boldsymbol{d}}_{\mathrm{tx},s})\), and \(\mathcal{L}_\mathrm{rx} = \mathcal{L}(\hat{\boldsymbol{r}}_{\mathrm{rx},p_2,k_2},{\boldsymbol{d}}_{\mathrm{rx},u})\) using \(\hat{\boldsymbol{r}}_{\mathrm{rx},p_2,k_2}=\boldsymbol{s}(\theta_{p_2,k_2,\mathrm{ZoA}},\phi_{p_2,k_2,\mathrm{AoA}})\). Furthermore, the \ac{Tx} Doppler term is \(\mathcal{D}_\mathrm{tx}(t) = \mathcal{D}(\hat{\boldsymbol{r}}_{\mathrm{tx},l,k},{\boldsymbol{v}}_{\mathrm{tx},l},t)\). The \ac{Rx} Doppler term, \(\mathcal{D}_\mathrm{rx}(t) = \mathcal{D}(\hat{\boldsymbol{r}}_{\mathrm{rx},p_2,k_2},{\boldsymbol{v}}_{\mathrm{rx},p_2},t)\), which employs \({\boldsymbol{v}}_{\mathrm{rx},p_2} = {\boldsymbol{v}}_\mathrm{rx}-{\boldsymbol{v}}_{p_2}\), capturing the motion between \ac{Rx} and deterministic clusters. The cluster Doppler term, \(\mathcal{D}_\mathrm{clus}(t) = \mathcal{D}(\hat{\boldsymbol{r}}_{l,k,p_2,k_2},{\boldsymbol{v}}_{l,p_2},t)\) incorporates \({\boldsymbol{v}}_{l,p_2} = {\boldsymbol{v}}_l-{\boldsymbol{v}}_{p_2}\), representing the relative motion between target and deterministic cluster in the target-\ac{Rx} link, where \(\hat{\boldsymbol{r}}_{l,k,p_2,k_2}=\boldsymbol{s}(\theta_{l,k,p_2,k_2,\mathrm{ZoD}},\phi_{l,k,p_2,k_2,\mathrm{AoD}})\). The deterministic cluster velocity is characterized by \({\boldsymbol{v}}_{p_2} = \boldsymbol{v}(v_{p_2},\theta_{p_2},\phi_{p_2})\) with speed \(v_{p_2}\) and direction angles \(\theta_{p_2}\), \(\phi_{p_2}\).
\paragraph{Case 3a: Stochastic \ac{NLoS} Component}

This scenario features a \ac{NLoS} \ac{Tx}-target link with stochastic clusters and a \ac{LoS} target-\ac{Rx} link, representing an instantiation of the generalized model in \eqref{cc_target} with \(\boldsymbol{\mathcal{Q}} = \{n_1,m_1,1\}\) and \(\zeta = \mathrm{SNLoS2,tar}\). The \ac{CC} integrate stochastic multipath components from the \ac{Tx}-target link with the \ac{LoS} component from the target-\ac{Rx} link. The complex channel gain is defined as \(\Gamma^{\mathrm{SNLoS2,tar}} = \sqrt{\sigma_{l,k} P_{n_1,m_1}}\), where \(P_{n_1,m_1}\) is the power of the \(m_1\)th ray within the \(n_1\)th stochastic cluster. Moreover, \(\boldsymbol{\Omega}_\mathrm{tx}^{\mathrm{SNLoS2,tar}}=\left\{\theta_{n_1,m_1,\mathrm{ZoD}},\phi_{n_1,m_1,\mathrm{AoD}}\right\}\) and \(\boldsymbol{\Omega}_\mathrm{rx}^{\mathrm{SNLoS2,tar}} = \left\{\theta_{l,k,\mathrm{ZoA}},\phi_{l,k,\mathrm{AoA}}\right\}\). The \ac{CPM} is given as \(\boldsymbol{\mathcal{C}}^{\mathrm{SNLoS2,tar}} = \boldsymbol{\mathcal{C}}(\kappa_{n_1,m_1},\boldsymbol{\Phi}_{n_1,m_1}^\mathrm{pol})\). Furthermore, \(\mathcal{L}_\mathrm{tx} = \mathcal{L}(\hat{\boldsymbol{r}}_{\mathrm{tx},n_1,m_1},{\boldsymbol{d}}_{\mathrm{tx},s})\), where \(\hat{\boldsymbol{r}}_{\mathrm{tx},n_1,m_1} = \boldsymbol{s}(\theta_{n_1,m_1,\mathrm{ZoD}},\phi_{n_1,m_1,\mathrm{AoD}})\), and \(\mathcal{L}_\mathrm{rx} = \mathcal{L}(\hat{\boldsymbol{r}}_{\mathrm{rx},l,k},{\boldsymbol{d}}_{\mathrm{rx},u})\). The \ac{Tx} Doppler term is \(\mathcal{D}_\mathrm{tx}(t) = \mathcal{D}(\hat{\boldsymbol{r}}_{\mathrm{tx},n_1,m_1},{\boldsymbol{v}}_\mathrm{tx},t)\) and the \ac{Rx} Doppler term is  \(\mathcal{D}_\mathrm{rx}(t) = \mathcal{D}(\hat{\boldsymbol{r}}_{\mathrm{rx},l,k},{\boldsymbol{v}}_{\mathrm{rx},l},t)\). The cluster Doppler term, \(\mathcal{D}_\mathrm{clus}(t) = \mathcal{D}(\hat{\boldsymbol{r}}_{l,n_1,m_1},{\boldsymbol{v}}_l,t)\) incorporates the target's motion relative to the cluster, where \(\hat{\boldsymbol{r}}_{l,n_1,m_1}=\boldsymbol{s}(\theta_{n_1,m_1,\mathrm{ZoA}},\phi_{n_1,m_1,\mathrm{AoA}})\).
\paragraph{Case 3b: Deterministic \ac{NLoS} Component}
This scenario features a \ac{NLoS} \ac{Tx}-target link due to deterministic clusters and a \ac{LoS} target-\ac{Rx} link, corresponding to the generalized model in eq. \eqref{cc_target} with \(\boldsymbol{\mathcal{Q}} = \{p_1,k_1,1\}\) and \(\zeta = \mathrm{DNLoS2,tar}\). The \ac{CC} combine multipath components from deterministic clusters in the \ac{Tx}-target link with the \ac{LoS} component in the target-\ac{Rx} link. In this propagation scenario, \(\Gamma^\mathrm{DNLoS2,tar} = \sqrt{\sigma_{l,k}\sigma_{p_1,k_1}P_{p_1,k_1}}\), where \(\sigma_{p_1,k_1}\) is the \ac{RCS} of the deterministic cluster and \(P_{p_1,k_1}\) is the power of a given \ac{SP} in the deterministic cluster. The angular parameters are \(\boldsymbol{\Omega}_\mathrm{tx}^{\mathrm{DNLoS2,tar}}=\left\{\theta_{p_1,k_1,\mathrm{ZoD}},\phi_{p_1,k_1,\mathrm{AoD}}\right\}\) and \(\boldsymbol{\Omega}_\mathrm{rx}^{\mathrm{DNLoS2,tar}} = \left\{\theta_{l,k,\mathrm{ZoA}},\phi_{l,k,\mathrm{AoA}}\right\}\). The \ac{CPM} for this propagation scenario is given as \(\boldsymbol{\mathcal{C}}^{\mathrm{DNLoS2,tar}} = \boldsymbol{\mathcal{C}}\left(\kappa_{p_1,k_1},\boldsymbol{\Phi}_{p_1,k_1}^\mathrm{pol}\right)\). The spatial location terms are \(\mathcal{L}_\mathrm{tx} = \mathcal{L}(\hat{\boldsymbol{r}}_{\mathrm{tx},p_1,k_1},{\boldsymbol{d}}_{\mathrm{tx},s})\) with \(\hat{\boldsymbol{r}}_{\mathrm{tx},p_1,k_1}=\boldsymbol{s}(\theta_{p_1,k_1,\mathrm{ZoD}},\phi_{p_1,k_1,\mathrm{AoD}})\), and \(\mathcal{L}_\mathrm{rx} = \mathcal{L}(\hat{\boldsymbol{r}}_{\mathrm{rx},l,k},{\boldsymbol{d}}_{\mathrm{rx},u})\), where \(\hat{\boldsymbol{r}}_{\mathrm{rx},l,k}=\boldsymbol{s}(\theta_{l,k,\mathrm{ZoA}},\phi_{l,k,\mathrm{AoA}})\). The \ac{Tx} Doppler term is \(\mathcal{D}_\mathrm{tx}(t) = \mathcal{D}(\hat{\boldsymbol{r}}_{\mathrm{tx},p_1,k_1},{\boldsymbol{v}}_{\mathrm{tx},p_1},t)\) which employs the relative velocity \({\boldsymbol{v}}_{\mathrm{tx},p_1} = {\boldsymbol{v}}_\mathrm{tx}-{\boldsymbol{v}}_{p_1}\) between \ac{Tx} and deterministic cluster, whereas, the \ac{Rx} Doppler term is \(\mathcal{D}_\mathrm{rx}(t) = \mathcal{D}(\hat{\boldsymbol{r}}_{\mathrm{rx},l,k},{\boldsymbol{v}}_{\mathrm{rx},l},t)\). The cluster Doppler term, \(\mathcal{D}_\mathrm{clus}(t) = \mathcal{D}(\hat{\boldsymbol{r}}_{l,k,p_1,k_1},{\boldsymbol{v}}_{l,p_1},t)\) incorporates \({\boldsymbol{v}}_{l,p_1} = {\boldsymbol{v}}_l-{\boldsymbol{v}}_{p_1}\), representing the relative motion between target and deterministic cluster, where \(\hat{\boldsymbol{r}}_{l,k,p_1,k_1}=\boldsymbol{s}(\theta_{l,k,p_1,k_1,\mathrm{ZoA}},\phi_{l,k,p_1,k_1,\mathrm{AoA}})\). The deterministic cluster velocity is characterized by \({\boldsymbol{v}}_{p_1} = \boldsymbol{v}(v_{p_1},\theta_{p_1},\phi_{p_1})\) with speed \(v_{p_1}\) and direction angles \(\theta_{p_1}\), \(\phi_{p_1}\).
\paragraph{Case 4a: Stochastic \ac{NLoS} Component}
 This scenario represents a propagation environment where both the \ac{Tx}-target and target-\ac{Rx} links experience \ac{NLoS} conditions due to stochastic clusters. Instantiated from the generalized model in eq. \eqref{cc_target} with \(\boldsymbol{\mathcal{Q}} = \{n_1,m_1,n_2,m_2\}\) and \(\zeta = \mathrm{SNLoS3,tar}\). The complex channel gain is \(\Gamma^{\mathrm{SNLoS3,tar}} = \sqrt{\sigma_{l,k}P_{n_1,m_1}P_{n_2,m_2}}\). The angular parameters are given as  \(\boldsymbol{\Omega}_\mathrm{tx}^{ \mathrm{SNLoS3,tar}}=\left\{\theta_{n_1,m_1,\mathrm{ZoD}},\phi_{n_1,m_1,\mathrm{AoD}}\right\}\) and \(\boldsymbol{\Omega}_\mathrm{rx}^{ \mathrm{SNLoS3,tar}} = \left\{\theta_{n_2,m_2,\mathrm{ZoA}},\phi_{n_2,m_2,\mathrm{AoA}}\right\}\). Moreover, the \ac{CPM} is given as \(\boldsymbol{\mathcal{C}}^{\mathrm{SNLoS3,tar}} = \boldsymbol{\mathcal{C}}(\kappa_{n_1,m_1,n_2,m_2},\boldsymbol{\Phi}_{n_1,m_1,n_2,m_2}^\mathrm{pol})\). The spatial location term for the \ac{Tx} is \(\mathcal{L}_\mathrm{tx} = \mathcal{L}(\hat{\boldsymbol{r}}_{\mathrm{tx},n_1,m_1},{\boldsymbol{d}}_{\mathrm{tx},s})\). Similarly, the \ac{Rx} spatial location term is \(\mathcal{L}_\mathrm{rx} = \mathcal{L}(\hat{\boldsymbol{r}}_{\mathrm{rx},n_2,m_2},{\boldsymbol{d}}_{\mathrm{rx},u})\). The \ac{Tx} and \ac{Rx} Doppler terms are \(\mathcal{D}_\mathrm{tx}(t)=\mathcal{D}\left(\hat{\boldsymbol{r}}_{\mathrm{tx},n_1,m_1},{\boldsymbol{v}}_\mathrm{tx},t\right)\) and \(\mathcal{D}_\mathrm{rx}(t)=\mathcal{D}\left(\hat{\boldsymbol{r}}_{\mathrm{rx},n_2,m_2},{\boldsymbol{v}}_\mathrm{rx},t\right)\), respectively. Moreover, the cluster Doppler terms is \(\mathcal{D}_\mathrm{clus}(t) = \mathcal{D}(\hat{\boldsymbol{r}}_{l,n_1,m_1},{\boldsymbol{v}}_l,t)\mathcal{D}(\hat{\boldsymbol{r}}_{l,n_2,m_2},{\boldsymbol{v}}_l,t)\).
\paragraph{Case 4b: Deterministic \ac{NLoS} Component}
This configuration represents the propagation scenario with double-bounce scattering through deterministic clusters in both \ac{Tx}-target and target-\ac{Rx} links. Instantiated from eq. \eqref{cc_target} with \(\boldsymbol{\mathcal{Q}} = \{p_1,k_1,p_2,k_2\}\) and \(\zeta = \mathrm{DNLoS3,tar}\). For this case, \(\Gamma^{\mathrm{DNLoS3,tar}} = \sqrt{\sigma_{l,k}\sigma_{p_1,k_1}\sigma_{p_2,k_2}P_{p_1,k_1}P_{p_2,k_2}}\). The angular components are given as \(\boldsymbol{\Omega}_\mathrm{tx}^{\mathrm{DNLoS3,tar}}=\left\{\theta_{p_1,k_1,\mathrm{ZoD}},\phi_{p_1,k_1,\mathrm{AoD}}\right\}\) and \(\boldsymbol{\Omega}_\mathrm{rx}^{\mathrm{DNLoS3,tar}} = \left\{\theta_{p_2,k_2,\mathrm{ZoA}},\phi_{p_2,k_2,\mathrm{AoA}}\right\}\). The \ac{CPM} is \(\boldsymbol{\mathcal{C}}^{\mathrm{DNLoS3,tar}} = \boldsymbol{\mathcal{C}}(\kappa_{p_1,k_1,p_2,k_2},\boldsymbol{\Phi}_{p_1,k_1,p_2,k_2}^\mathrm{pol})\). The spatial location terms for the \ac{Tx} and the \ac{Rx} are given as \(\mathcal{L}_\mathrm{tx} = \mathcal{L}(\hat{\boldsymbol{r}}_{\mathrm{tx},p_1,k_1},{\boldsymbol{d}}_{\mathrm{tx},s})\) and \(\mathcal{L}_\mathrm{rx} = \mathcal{L}(\hat{\boldsymbol{r}}_{\mathrm{rx},p_2,k_2},{\boldsymbol{d}}_{\mathrm{rx},u})\), respectively. Lastly, the \ac{Tx} and \ac{Rx} Doppler terms are given as \(\mathcal{D}_\mathrm{tx}(t)=\mathcal{D}\left(\hat{\boldsymbol{r}}_{\mathrm{tx},p_1,k_1},{\boldsymbol{v}}_{\mathrm{tx},p_1},t\right)\) and  \(\mathcal{D}_\mathrm{rx}(t)=\mathcal{D}\left(\hat{\boldsymbol{r}}_{\mathrm{rx},p_2,k_2},{\boldsymbol{v}}_{\mathrm{rx},p_2},t\right)\), respectively. Furthermore, the cluster Doppler term is \(\mathcal{D}_\mathrm{clus}(t)=\mathcal{D}(\hat{\boldsymbol{r}}_{\mathrm{tx},l,k,p_1,k_1},{\boldsymbol{v}}_{l,p_1},t)\mathcal{D}(\hat{\boldsymbol{r}}_{l,k,p_2,k_2},{\boldsymbol{v}}_{l,p_2},t)\).

The composite \ac{CIR} formulations integrate both stochastic and deterministic cluster components across the different propagation scenarios. A generalized expression to obtain the \ac{NLoS} \ac{CIR} for the above-mentioned cases is given as:
\begin{equation}\label{cir_NLOS_general}  
\begin{split}
H_{u,s,l,\boldsymbol{\mathcal{N}},\boldsymbol{\mathcal{P}}}^{\mathrm{NLoS}\xi,\mathrm{tar}}(t,\tau) = \sum_{k} &\left(\sum_{\boldsymbol{m} \in \mathcal{M}_\xi} H_{u,s,\boldsymbol{\mathcal{Q}}}^\mathrm{SNLoS\xi,tar}(t)\delta(\tau - \tau_\xi^S)\right. \\
&~\left.+\sum_{\boldsymbol{k} \in \mathcal{K}_\xi} H_{u,s,\boldsymbol{\mathcal{Q}}}^\mathrm{DNLoS\xi,tar}(t)\delta(\tau - \tau_\xi^D) \right)
\end{split}
\end{equation}
where \(\xi \in \{1,2,3\}\) denotes the case index, \(\boldsymbol{\mathcal{N}}\) represents the set of stochastic cluster indices, \(\boldsymbol{\mathcal{P}}\) represents the set of deterministic cluster indices, \(\boldsymbol{\mathcal{Q}}\) denotes the ray/cluster index vector, \(\mathcal{M}_\xi\) and \(\mathcal{K}_\xi\) are the summation domains for stochastic and deterministic components in case \(\xi\), and \(\tau_\xi^S\), \(\tau_\xi^D\) are the corresponding stochastic and deterministic path delays.

The generalized \ac{CIR} formulation in eq. \eqref{cir_NLOS_general} is instantiated for each specific propagation case through appropriate parameter mappings. For Case 1 (\(\xi = 1\)), corresponding to \ac{LoS} \ac{Tx}-target and \ac{NLoS} target-\ac{Rx} propagation, the parameters are specified as \(\boldsymbol{\mathcal{N}} = \{n_2\}\), \(\boldsymbol{\mathcal{P}} = \{p_2\}\), \(\boldsymbol{\mathcal{Q}} = \{1,n_2,m_2\}\), \(\mathcal{M}_1 = \{m_2\}\), \(\mathcal{K}_1 = \{k_2\}\), with delay parameters \(\tau_1^S = \tau_1\) and \(\tau_1^D = \tau_2\). Here, \(\tau_1 = \tau_{l,s} - \tau_{n_2,m_2}\) represents the net delay through the target and stochastic cluster path, and \(\tau_2 = \tau_{l,k,s} - \tau_{u,l,k,p_2,k_2}\) captures the deterministic path delay via the target's \ac{SP} and deterministic cluster. The constituent delays include \(\tau_{l,s}\) denoting the propagation delay from the \(s\)th transmit antenna to the \(l\)th target, \(\tau_{n_2,m_2}\) as the delay associated with the \(m_2\)th ray in the \(n_2\)th stochastic cluster, and \(\tau_{u,l,k,p_2,k_2} = (\Vert {\boldsymbol{d}}_{p_2,k_2}-{\boldsymbol{d}}_{l,k}\Vert + \Vert {\boldsymbol{d}}_{\mathrm{rx},u}-{\boldsymbol{d}}_{p_2,k_2}\Vert)/c\) representing the bistatic delay through the deterministic cluster path. For Case 2 (\(\xi = 2\)), representing \ac{NLoS} \ac{Tx}-target and \ac{LoS} target-\ac{Rx} propagation, the parameter mappings are \(\boldsymbol{\mathcal{N}} = \{n_1\}\), \(\boldsymbol{\mathcal{P}} = \{p_1\}\), \(\boldsymbol{\mathcal{Q}} = \{n_1,m_1,1\}\), \(\mathcal{M}_2 = \{m_1\}\), \(\mathcal{K}_2 = \{k_1\}\), with delays \(\tau_2^S = \tau_3\) and \(\tau_2^D = \tau_4\). The delay parameters are defined as \(\tau_3 = \tau_{n_1,m_1} - \tau_{u,l}\) for the stochastic path and \(\tau_4 = \tau_{s,l,k,p_1,k_1} - \tau_{u,l,k}\) for the deterministic path, where \(\tau_{u,l}\) is the target to \ac{Rx} delay, \(\tau_{n_1,m_1}\) is the stochastic cluster ray delay, \(\tau_{s,l,k,p_1,k_1} = (\Vert {\boldsymbol{d}}_{p_1,k_1}-{\boldsymbol{d}}_{\mathrm{tx},s}\Vert + \Vert {\boldsymbol{d}}_{l,k}-{\boldsymbol{d}}_{p_1,k_1}\Vert)/c\) represents the \ac{Tx} to target delay via deterministic cluster, and \(\tau_{u,l,k} = \Vert {\boldsymbol{d}}_{\mathrm{rx},u}-{\boldsymbol{d}}_{l,k} \Vert/c\) is the direct target-\ac{Rx} path delay. Case 3 (\(\xi = 3\)) characterizes \ac{NLoS} propagation in both links and employs the parameter set \(\boldsymbol{\mathcal{N}} = \{n_1,n_2\}\), \(\boldsymbol{\mathcal{P}} = \{p_1,p_2\}\), \(\boldsymbol{\mathcal{Q}} = \{n_1,m_1,n_2,m_2\}\), \(\mathcal{M}_3 = \{(m_1,m_2)\}\), \(\mathcal{K}_3 = \{k_1,k_2\}\), with delays \(\tau_3^S = \tau_5\) and \(\tau_3^D = \tau_6\), where \(\tau_5 = \tau_{n_1,m_1} - \tau_{n_2,m_2}\) represents the stochastic-stochastic path delay and \(\tau_6 = \tau_{s,l,k,p_1,k_1} - \tau_{u,l,k,p_2,k_2}\) captures the deterministic-deterministic path delay.

The vector \ac{CIR} components are defined as \(
\mathbf{H}_{u,s}^\mathrm{LoS,tar}(t,\tau) = \left[H_{u,s,l,1,1}^\mathrm{LoS,tar}(t)\delta(\tau-\tau_{u,s,l})\right]\forall~l\), with \(H_{u,s,l,1,1}^\mathrm{LoS,tar}(t) = \sum_k H_{u,s,l,k,1,1}^\mathrm{LoS,tar}(t)\), \(\mathbf{H}_{u,s}^\mathrm{NLoS1, tar}(t,\tau) = \left[H_{u,s,l,n_2,p_2}^\mathrm{NLoS1, tar}(t)\delta(\tau-\tau_1)\right]\forall~l,n_2,p_2\), \(
\mathbf{H}_{u,s}^\mathrm{NLoS2, tar}(t,\tau) = \left[H_{u,s,l,n_1,p_1}^\mathrm{NLoS2, tar}(t)\delta(\tau-\tau_3)\right]\forall~l,n_1,p_1\), and \(
\mathbf{H}_{u,s}^\mathrm{NLoS3, tar}(t,\tau) = \left[H_{u,s,l,n_1,n_2,p_1,p_2}^\mathrm{NLoS3, tar}(t)\delta(\tau-\tau_5)\right]\forall~l,n_1,p_1,n_2,p_2\), where each vector represents the \ac{CIR} across discrete delay bins, with non-zero elements only at the specific delay taps corresponding to their respective propagation paths.

The target \ac{CIR} is characterized by distinct formulations determined by the propagation conditions of the constituent \ac{Tx}-target and target-\ac{Rx} links. A unified expression encompassing all propagation scenarios is given by:
\begin{equation}\label{target_cir_comb}
    \begin{split}
        \boldsymbol{\mathcal{H}}_{u,s}^\mathrm{tar}(t,\tau) =& \gamma_1 \mathbf{H}_{u,s}^\mathrm{LoS,tar}(t,\tau) + \gamma_2\mathbf{H}_{u,s}^\mathrm{NLoS1, tar}(t,\tau) \\
        &+ \gamma_3\mathbf{H}_{u,s}^\mathrm{NLoS2, tar}(t,\tau) + \gamma_4 \boldsymbol{\mathcal{H}}_{u,s}^\mathrm{NLoS3, tar}(t,\tau),
    \end{split}
\end{equation}
where the weighting coefficients \(\gamma_i\) are configured according to the specific link conditions. The coefficients are derived from the Rician \(K\)-factors of the individual links: \(\eta_1 = \sqrt{K_1^\mathrm{tar}/(1+K_1^\mathrm{tar})}\) and \(\eta_2 = \sqrt{K_2^\mathrm{tar}/(1+K_2^\mathrm{tar})}\) represent the specular component weights for the \ac{Tx}-target and target-\ac{Rx} links respectively, while \(\tilde{\eta}_1 = \sqrt{1/(1+K_1^\mathrm{tar})}\) and \(\tilde{\eta}_2 = \sqrt{1/(1+K_2^\mathrm{tar})}\) correspond to the diffuse component weights.

For \ac{LoS} propagation, the coefficients are defined as \(\gamma_1 = \eta_1\eta_2\), \(\gamma_2 = \eta_1\tilde{\eta}_2\), \(\gamma_3 = \tilde{\eta}_1\eta_2\), and \(\gamma_4 = \tilde{\eta}_1\tilde{\eta}_2\). When the \ac{Tx}-target link maintains \ac{LoS} conditions while the target-\ac{Rx} link transitions to \ac{NLoS} propagation, the coefficients are configured as \(\gamma_1 = 0\), \(\gamma_2 = \eta_1\), \(\gamma_3 = 0\), and \(\gamma_4 = \tilde{\eta}_2\). Conversely, for \ac{NLoS} propagation in the \ac{Tx}-target link combined with \ac{LoS} conditions in the target-\ac{Rx} link, the coefficients become \(\gamma_1 = 0\), \(\gamma_2 = 0\), \(\gamma_3 = \eta_2\), and \(\gamma_4 = \tilde{\eta}_1\). Finally, under \ac{NLoS} conditions in both links, the channel is characterized by \(\gamma_1 = 0\), \(\gamma_2 = 0\), \(\gamma_3 = 0\), and \(\gamma_4 = 1\).
\begin{algorithm*}[!t]
\small
\caption{Proposed Bistatic \ac{ISAC} Channel Generation Process}
\label{alg:isac_concise_math}
\begin{algorithmic}[1]
\STATE \textbf{Input:} Scenario (UMa/UMi/InF), $f_c$, Tx/Rx array geometries $ \mathbf{d}_{\mathrm{tx},s}$, $\mathbf{d}_{\mathrm{rx},u}$, 
Tx/Rx trajectories and velocities $\mathbf{v}_{\mathrm{tx}}$, $\mathbf{v}_{\mathrm{rx}}$, Targets $\{(\sigma_{l,k},\mathbf{d}_{l,k},\mathbf{v}_l)\}_{l=1,\cdots,L,k=1,\cdots,K}$, Deterministic clusters/\acp{EO} (Type-I) per link $\{(\sigma_{p_i,k_i},\mathbf{v}_{p_i})\}$; \ac{TR}38.901 parameters and functions (\ac{LSP}/\ac{SSP} generation, intra-cluster delay/ray generation, \ac{XPR}, phases).
\STATE \textbf{Output:} $\boldsymbol{\mathcal{H}}^{\mathrm{ISAC}}_{u,s}(t,\tau)$ for all antenna pairs $(u,s)$ and snapshots $t$.
\STATE \textbf{(I) General Parameters (Steps 1C-4C, 1S-4S):}
\STATE Set scenario/layout/coordinate systems and target/\ac{EO} parameters.
\STATE Determine \ac{LoS}/\ac{NLoS} states and \ac{LoS} angles for background and target segments (\ac{Tx}-target, target-\ac{Rx}).
\STATE Generate correlated \acp{LSP} (\ac{DS}/\ac{AS}/\ac{SF}/\(K\)) per \ac{TR}38.901 for each link.
\STATE \textbf{(II) Small-Scale Parameters: (Steps 5C-9C, 5S-9S):}
\STATE \textbf{(A) Background link (Tx-Rx):}
\STATE Generate $\tau'_n$ and set baseline $\tau_{\mathrm{LoS}}=d_{\mathrm{Tx,Rx}}/c$.
\STATE Compute absolute delays $\tau_0=\tau_{\mathrm{LoS}}$, and for $n>0$: $\tau_n=\tau'_n+\tau_{\mathrm{LoS}}+\Delta\tau$; generate ray delays $\tau_{n,m}$ via intra-cluster procedure. \label{line:bg_abs_delay}
\STATE Generate $P_{n,m}$, $\mathbf{\Omega}^{\zeta}_{\mathrm{tx}},\mathbf{\Omega}^{\zeta}_{\mathrm{rx}}$, random ray coupling, and $\kappa_{n,m}$.
\vspace{0.3em}
\STATE \textbf{(B) Target links (\ac{Tx}-target and target-\ac{Rx}):}
\STATE Generate $\tau'_{n_1}$ and $\tau'_{n_2}$ for both links; each has its own baseline 
$\tau^{(1)}_{\mathrm{LoS}}=d_{\mathrm{Tx,tar}}/c$ and $\tau^{(2)}_{\mathrm{LoS}}=d_{\mathrm{tar,Rx}}/c$.\label{line:seg_baselines}
\STATE Partition \ac{NLoS} clusters into $(D_1,S_1)$ for \ac{Tx}-target and $(D_2,S_2)$ for target-\ac{Rx}.
\label{line:partition}
\STATE Compute absolute delays: for \(S_1\) and \(S_2\): $\tau_{n_1}=\tau'_{n_1}+\tau^{(1)}_{\mathrm{LoS}}+\Delta\tau^{(1)}$, \quad $\tau_{n_2}=\tau'_{n_2}+\tau^{(2)}_{\mathrm{LoS}}+\Delta\tau^{(2)}$; for \(D_1\) and \(D_2\), $\tau_{p_1}=\tau'_{p_1}+\tau^{(1)}_{\mathrm{LoS}}$, \quad $\tau_{p_2}=\tau'_{p_2}+\tau^{(2)}_{\mathrm{LoS}}$.
\label{line:tar_abs_delay}
\STATE Generate ray/\ac{SP} delays $\tau_{n_i,m_i}$, $\tau_{p_i,k_i}$, powers, and \(\mathbf{\Omega}_\mathrm{tx}^\zeta\), \(\mathbf{\Omega}_\mathrm{rx}^\zeta\).
\STATE \textbf{(C) Spatial mapping of deterministic clusters (Step 7Sa):}
\FOR{each deterministic cluster $p_1$ (\ac{Tx}-target link) and \(p_2\) (target-\ac{Rx} link)}
  \STATE Let $\boldsymbol{r}_1$ be the \ac{Tx} position and $\boldsymbol{r}_2$ be the target position; define $\boldsymbol{d}=\boldsymbol{r}_2-\boldsymbol{r}_1$, $d=\|\boldsymbol{d}\|$.
  \STATE From the chosen angle set (\ac{AA} or \ac{AD}), form unit direction $\hat{\boldsymbol{a}}_{p}$ and set $d_p=c\,\tau_p$.
  \STATE Compute cluster range \(\boldsymbol{r}_p=\boldsymbol{r}_2+\hat{\boldsymbol{a}}_p\,|\boldsymbol{a}_p|\) with \( |\boldsymbol{a}_p|=\frac{d_p^2-d^2}{2\!\left(d_p-\boldsymbol{d}^{T}\hat{\boldsymbol{a}}_p\right)} \).
  \STATE Update complementary angles (\ac{AA}/\ac{AD}) from $\boldsymbol{r}_p$ to enforce spatiotemporal consistency (Step 7Sa).
\ENDFOR
\STATE (Optional) Apply the same mapping to infer target position for stochastic target placement.
\STATE \textbf{(D) Couple \ac{Tx}-target and target-\ac{Rx} links at the target (Step 8sb):}
\STATE Perform random ray coupling for stochastic clusters within each segment.
\STATE Enumerate all multipath combinations at the target: \ac{LoS}-\ac{LoS}: $\boldsymbol{\mathcal{Q}}=\{1,1\}$, \ac{LoS}-\ac{NLoS}: $\boldsymbol{\mathcal{Q}}=\{1,n_2,m_2\}$ and $\boldsymbol{\mathcal{Q}}=\{1,p_2,k_2\}$ \ac{NLoS}-\ac{LoS}: $\boldsymbol{\mathcal{Q}}=\{n_1,m_1,1\}$ and $\boldsymbol{\mathcal{Q}}=\{p_1,k_1,1\}$, and \ac{NLoS}-\ac{NLoS}: $\boldsymbol{\mathcal{Q}}=\{n_1,m_1,n_2,m_2\}$ and $\boldsymbol{\mathcal{Q}}=\{p_1,k_1,p_2,k_2\}.$ \label{line:coupling_cases}
\STATE Generate $\kappa$ for all coupled cases (stochastic and deterministic), with deterministic \ac{XPR} optionally incorporating \ac{EO} geometry/materials.

\STATE \textbf{(III) Coefficient Generation (Steps 10C-12C,10S-12S, 13 ):}
\STATE Draw initial phases for background and target channels (with the case-dependent indexing sets).
\FOR{each snapshot $t$ and antenna pair $(u,s)$}
    \STATE \textbf{(A) Background coefficients:}
    compute $H^{\zeta}_{u,s,\boldsymbol{\mathcal{Q}}}(t)$ using \eqref{cc_background}, for $\zeta\in\{\mathrm{LoS,back},\mathrm{NLoS,back}\}$ and the appropriate $\boldsymbol{\mathcal{Q}}$; then form $\boldsymbol{\mathcal{H}}^{\mathrm{back}}_{u,s}(t,\tau)$ according to \eqref{LOS_cir_background}. \label{line:bg_coeff}

    \STATE \textbf{(B) Target coefficients:} 
    compute $H^{\zeta}_{u,s,\boldsymbol{\mathcal{Q}}}(t)$ using \eqref{cc_target} for each coupled case (\ac{LoS}, stochastic-\ac{NLoS}, and deterministic-\ac{NLoS}).\label{line:tar_coeff}

    \STATE Assemble the \ac{NLoS} target \ac{CIR} using the composite formulation in \eqref{cir_NLOS_general}, and then combine the different cases through \eqref{target_cir_comb} to obtain $\boldsymbol{\mathcal{H}}^{\mathrm{tar}}_{u,s}(t,\tau)$.\label{line:tar_assemble}

    \STATE \textbf{(C) \ac{ISAC} superposition:} 
    obtain the overall \ac{ISAC} \ac{CIR} using \eqref{ISAC_chan}.
\ENDFOR
\end{algorithmic}
\end{algorithm*}
\section{Results, Analysis and Discussion}\label{sec4}
In this section, we present a comprehensive evaluation of the proposed \ac{ISAC} channel, focusing on both communication and sensing performance metrics. The simulations are conducted in \ac{3GPP} standardized \ac{UMi}, \ac{UMa}, and \ac{InF} environments. First, we analyze the communication performance by examining the \ac{BER} of the proposed \ac{ISAC} channel and comparing it with the classical \ac{TR}38.901 \cite{tr} {communication} channel model. Additionally, we compare the \ac{BER} performance in simulated and measured \ac{ISAC} channels. We also investigate the ergodic channel  and analyze the impact of targets with varying \ac{RCS}. For the sensing performance evaluation, we apply range estimation \cite{salmane2025high} and detection algorithms \cite{10681445} and assess their performance through the mean target range error and ROC curves, thereby linking the model to end-to-end detection probability under different target RCS values and distances. Note that the performance of other range estimation and detection algorithms might be different. Unless otherwise explicitly mentioned, the simulation parameters provided in Table \ref{sim_params} shall be used for the simulations. 
\begin{table}[tb]
    \centering
    \caption{Parameters used for simulations. In the table, following acronyms are used \ac{OFDM}, \ac{QAM}, and \ac{URA}.}
\footnotesize
\begin{tabular}{|l|l|}
        \hline
        \rowcolor{gray!20}
        \textbf{Category} & \textbf{Value/Setting} \\
        \hline
        \multicolumn{2}{|c|}{\textbf{System Configuration}} \\
        \hline
        Signaling & \ac{OFDM} \\
        Modulation Scheme & \(4\)-\ac{QAM} \\
        Number of Subcarriers (\(N\)) & \(256\) \\
        Length of CP & \(32\) \\
        Operating Frequency (\(f_c\)) & \SI{7}{GHz}\\
        Sampling Frequency (\(F_s\))& \SI{1}{GHz}  \\
        \hline
         \multicolumn{2}{|c|}{\textbf{Channel \& Environment} } \\
        \hline
        Channel Model & Proposed ISAC \\
        {}& \ac{TR}38.901\\
        Scenarios & \ac{UMi}, \ac{UMa}, \ac{InF} \\
        No. of \ac{UMi}/\ac{UMa} Clusters & \(12\) \\
        No. of \ac{InF} Clusters & \(10\) \\
        No. of Rays per Cluster & \(20\) \\
        No. of Deterministic Clusters & \(5\)\\
        No. of \acp{SP} of Deterministic Clusters & \(5\) \\
        \hline
         \multicolumn{2}{|c|}{\textbf{Antenna \& Mobility} } \\
        \hline
        \ac{Tx} Configuration & $2\times2$ URA \(\Rightarrow N_t =4\) \\
        \ac{Rx} Configuration & $2\times2$ URA \(\Rightarrow N_r =4\) \\
        \ac{Tx} Center Coordinates & \((0,0,5)\) m \\
        \ac{Rx} Center Coordinates & \((0,5,5)\) m \\
        \ac{Tx} Velocity & \(0\) m/s \\
        \ac{Rx} Velocity & \(0\) m/s \\
        \hline
         \multicolumn{2}{|c|}{\textbf{Target, Cluster \& Clutter Properties} } \\
        \hline
        No. of Targets & \(1\) (Figs. \ref{chan_cap} and \ref{rocs})\\
        {} & \(2\) (Fig. \ref{pos_error})\\
        Target Coordinates & Target 1:\((3,2,5)\) m \\
        {}& Target 2:\((10,5,1.5)\) m \\
        Target Velocity & \(0\) m/s \\
        \ac{RCS} of Target & Varied \\
        No. of \acp{SP} of Target & \(5\)\\
       \ac{RCS} of Deterministic Clusters & \(0.1\)m$^2$\\
        Deterministic Cluster Velocity & \(0\) m/s \\
        \ac{InF} Clutter Density & \(10\%\) \\
        \ac{InF} Clutter Height & \(2\) m \\
        \hline
    \end{tabular}
    \label{sim_params}
\end{table}
\subsection{Communication Performance Evaluation}\label{sec4a}
In this section, we evaluate the communication performance for the proposed \ac{ISAC} channel model, which is benchmarked against the standardized \ac{TR}38.901 channel \cite{tr}.
\subsubsection{BER Analysis in \ac{UMi}, \ac{UMa}, and \ac{InF} Scenarios}
\begin{figure}[tb]
    \centering
    \includegraphics[trim={12mm 1mm 1mm 0mm},clip,width=0.45\textwidth]{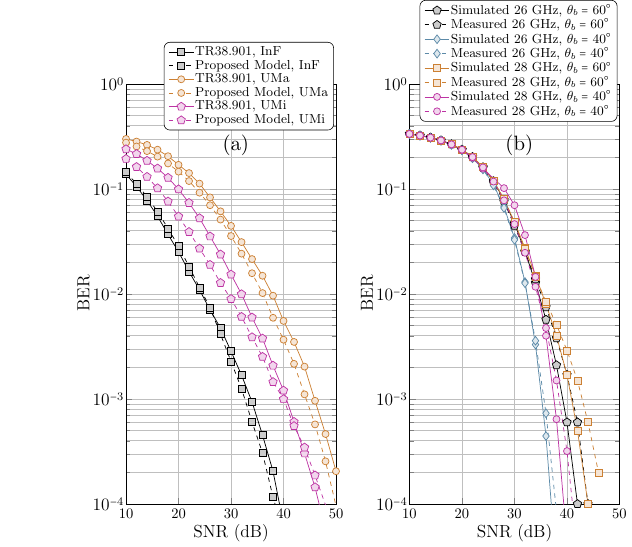} 
    \caption{\ac{BER} performance comparison: (a) proposed \ac{ISAC} model versus standard \ac{TR}38.901 \cite{tr} in \ac{UMi}, \ac{UMa}, and \ac{InF} scenarios with a single target (\ac{RCS} = \SI{0.1}{\meter^2}), (b) proposed model versus measured \ac{ISAC} channel in an \ac{InF} scenario under the same target conditions.}
    \label{ber_eniv} 
\end{figure}
Fig. \ref{ber_eniv}(a) presents a comparative analysis of the \ac{BER} performance between the proposed bistatic \ac{ISAC} channel model and the conventional \ac{TR}38.901 \cite{tr} channel model. The results {obtained considering a target \ac{RCS} of \SI{0.1}{\meter^2} demonstrate} near-equivalent performance between both channels, with the proposed \ac{ISAC} channel exhibiting marginally superior \ac{BER}. It can be observed that to achieve a \ac{BER} of \(10^{-3}\), the \ac{SNR} needed in case of classical \ac{TR}38.901 \ac{InF} is \(35\) dB, whereas, for the proposed model \ac{InF}, the \ac{SNR} needed is approximately \(34\) dB. Similarly, for \ac{UMa} and {UMi} of the \ac{TR}38.901, the \ac{SNR} required to achieve a \ac{BER} of \(10^{-3}\) is approximately \(47.5\) dB and \(40.5\) dB, respectively. For the same \ac{BER}, the proposed model \ac{UMa} and \ac{UMi} scenarios require \ac{SNR} pf \(46\)dB and \(40\) dB, respectively. This performance enhancement can be attributed to two key factors: (1) the incorporation of target with multiple \acp{SP} that introduce additional multipath components, and (2) the presence of deterministic clusters with defined \ac{RCS} values, as opposed to purely stochastic clusters in \ac{TR}38.901 \cite{tr} {communications model}. These deterministic clusters provide more structured reflection paths, thereby improving signal reliability. The close alignment in BER performance across different environments affirms that the proposed ISAC channel exhibits similar \ac{BER} as existing communication standards while simultaneously enabling sensing functionality. Furthermore, the model's adherence to 3GPP parameters ensures seamless integration with current 5G-\ac{NR} and future 6G systems, making it particularly suitable for applications requiring simultaneous communication and sensing.
\subsubsection{Comparative BER Performance Analysis: Simulated vs. Measured ISAC Channels}
In this section, we present a comparative analysis of the \ac{BER} performance between the simulated and measured \ac{ISAC} channels under two distinct bistatic configurations: \(40^\circ\) and \(60^\circ\) as shown in Fig. \ref{ber_eniv}(b). The performance comparison is constrained to a \ac{SISO}. For the target characterization, the simulation employs statistical \ac{RCS} modeling based on empirical measurements \cite{azim2025rcs}. The \ac{RCS} of the \ac{UAV} (small-sized) when modeled using log-normal distribution according to \ac{3GPP} specifications gives \(A = -13.57\)dBsm, \(B_1 = 0\)dB, \(B_2 = 3.065\)dB \cite{azim2025rcs}. The experimental measurements utilize a DJI Mavic \(2\) Pro \ac{UAV} with folded dimensions of \(214\times 91 \times 84\)mm, unfolded dimensions of \(322\times242\times84\)mm, a mass of \(907\)g, and a construction comprising magnesium alloy, reinforced plastic, carbon fiber, glass, and silicon components \cite{azim2025rcs}. The simulated channel considers clutter density of \(10\%\) and a clutter height of \SI{2}{\meter} to emulate \ac{InF} propagation conditions.
\begin{figure}[tb]
    \centering
    \includegraphics[trim={1mm 1mm 1mm 0mm},clip,width=0.43\textwidth]{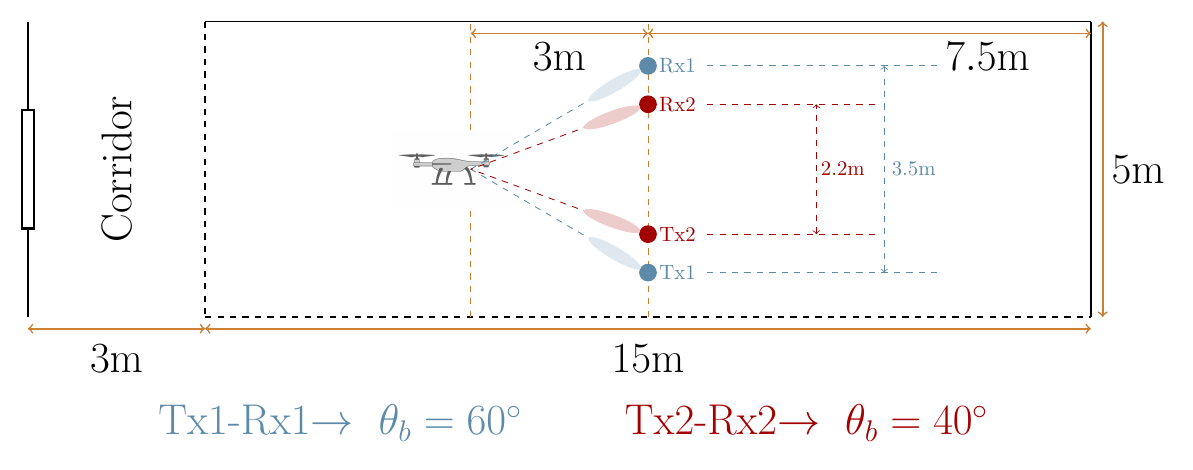} 
    \caption{The layout of the InF measurement environment depicting the \ac{Tx}-\ac{Rx} separation for \(40^\circ\) and \(60^\circ\) bistatic configurations. }
    \label{layout} 
\end{figure}

\begin{figure}[tb]
    \centering
    \includegraphics[trim={1mm 1mm 1mm 0mm},clip,width=0.49\textwidth]{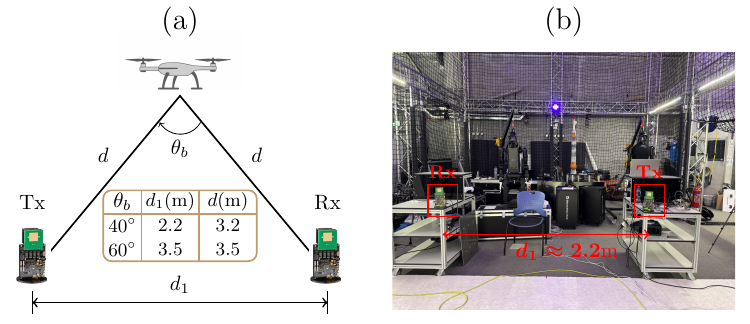} 
    \caption{(a) Bistatic measurement setup configuration, (b) actual measurement setup depicting \(40^\circ\) bistatic arrangement for the \ac{Tx} and the \ac{Rx}.}
    \label{setup} 
\end{figure}

The experimental measurements were conducted in the KINESIS Lab at NYU Abu Dhabi's Core Technology Platforms, which emulates \ac{InF} propagation characteristics. The environment measures \SI{5}{\meter} (width) \(\times\) \SI{15}{\meter} (length) \(\times\) \SI{8.5}{\meter} (height) and incorporates structural features typical of \ac{InF} environments, such as high ceilings and open spatial configurations. The experimental setup {layout}, including the placement of the \ac{Tx}, \ac{Rx} is illustrated in Fig.~\ref{layout} for both bistatic configurations. To ensure consistency between the simulated and experimental environments, the simulation parameters including the room dimensions and the \ac{Tx}, \ac{Rx}, and target coordinates were carefully aligned with the measurement setup. The origin \((0,0,0)\)m was defined at the center of the measurement environment (excluding the corridor), with all positions referenced accordingly. For the \(40^\circ\) bistatic configuration, the \ac{Tx}, \ac{Rx}, and target were positioned at \((0,-1.1,1)\)m, \((0,1.1,1)\)m, and \((3,0,1\))m, respectively. In the \(60^\circ\) case, the coordinates were adjusted to \((0,-1.75,1)\)m (\ac{Tx}), \((0,1.75,1)\)m (\ac{Rx}), and \((3,0,1)\)m (target). The carrier frequencies are set to \SI{26}{\GHz} and \SI{28}{\GHz} and the bandwidth limited to \SI{20}{\MHz} due to the \ac{USRP} constraints. Moreover, we employ a \SI{2.2}{\meter} and \SI{3.5}{\meter} \ac{Tx}-\ac{Rx} baseline separation $d_1$, resulting in bistatic angle of \SI{40}{\degree} and \SI{60}{\degree} respectively, at the target coordinates. Both \ac{Tx}-target and \ac{Rx}-target distances $d$ are equal to \SI{3.2}{\meter} and \SI{3.5}{\meter} for \SI{40}{\degree} and \SI{60}{\degree} of bistatic angles respectively as shown in Fig.~\ref{layout}. {The bistatic measurement setup configuration and the actual measurement setup indicating a \SI{40}{\degree} bistatic angle is illustrated in Fig. \ref{setup}(a) and Fig. \ref{setup}(b), respectively.}

The \ac{Tx} uses MATLAB-generated Zadoff-Chu sequences of length \(128\) for waveform generation, upconverted to the \SI{26}{\GHz} and \SI{28}{\GHz} via a Sivers EVK02004 \ac{RFIC} interfaced with a B205mini \ac{USRP}. The \ac{Rx} chain performs downconversion and digitization using an identical \ac{USRP}-\ac{RFIC} pair. An intermediate frequency of \SI{3}{\GHz} is used at both the \ac{Tx} and the \ac{Rx}. During the measurements, both \ac{Tx} and \ac{Rx} antennas bore-sight pattern (HPBW = \SI{10}{\degree} in zenith/azimuth) is carefully aligned to converge at the target's position. It is highlighted that the \ac{UAV} was rotating around a fixed position to inherently capture the angular effects on the received \ac{CIR}. Real-time processing is performed to extract \ac{CIR} from the received waveforms. 
This experimentally obtained \ac{CIR} is then employed in simulations where an \ac{OFDM} waveform (with parameters defined in Table \ref{sim_params}) is used for \ac{BER} performance evaluation. All the measurements were conducted in \ac{LoS} propagation conditions with horizontal polarization for both the \ac{Tx} and the \ac{Rx}. The simulated \ac{CIR} of the proposed \ac{ISAC} channel is utilized to assess theoretical \ac{BER} performance.

A comparative analysis of the simulated and measured \ac{BER} performance at carrier frequencies of \SI{26}{GHz} and \SI{28}{GHz}, under bistatic angles of \(40^\circ\) and \(60^\circ\), is presented in Fig. \ref{ber_eniv}(b). The results indicate a consistent behavior across all configurations, wherein the measured \ac{ISAC} channel exhibits marginally elevated \ac{BER} relative to its simulated counterpart at high \ac{SNR} values. For a \ac{BER} of \(10^{-3}\), the required \ac{SNR} was evaluated using both simulated and measured channel models. At a carrier frequency of \SI{26}{\giga\hertz}, the simulated channel requires an \ac{SNR} of approximately \SI{35}{dB} and \SI{40.75}{dB} for bistatic angles of \ang{40} and \ang{60}, respectively. In contrast, when using the measured channel, the required \acp{SNR} are \SI{35.5}{dB} and \SI{37.7}{dB} for the same angles. At \SI{28}{\giga\hertz}, the required \ac{SNR} for the simulated channel is \SI{37}{dB} and \SI{39.5}{dB} for bistatic angles of \ang{40} and \ang{60}. The corresponding values for the measured channel at this frequency are \SI{37.7}{dB} and \SI{40.75}{dB}. The \ac{BER} gap obtained using the simulated and measured channels in Fig. \ref{ber_eniv}(b) is due to cumulative model-measurement mismatch. First, the simulation assumes an idealized link-level \ac{Rx} operating on the modeled/extracted \ac{CIR}, whereas the measurement chain is affected by practical radio-frequency front-end impairments, including residual carrier-frequency and phase offsets, phase noise, IQ imbalance, finite synchronization accuracy, and imperfect amplitude/phase calibration. Second, the target response in simulation is represented through a statistical \ac{RCS} model, while the measured \ac{UAV} exhibits aspect- and material-dependent scattering that varies with orientation. Since the \ac{UAV} was rotated during the measurements to capture angular dependence, the effective target scattering response becomes time-varying and could deviate from the assumed statistical abstraction. Third, the measurement environment may contain weak site-specific propagation components, clutter, antenna misalignment effects, and other parasitic reflections that are not fully represented in the geometric/statistical simulation model. Therefore, the slightly higher \ac{BER} observed for the measured channel at high \ac{SNR} is attributed to the aggregate effect of residual hardware impairments, calibration uncertainty, target-\ac{RCS} mismatch, and unmodeled environmental propagation mechanisms.
\subsubsection{Ergodic Channel Capacity and Impact of RCS}

In this section, we perform a comparative analysis of the ergodic channel capacity across standardized \ac{UMi}, \ac{UMa}, and \ac{InF} environments considering the proposed \ac{ISAC} channel and compare it with the standardized \ac{TR}38.901 \cite{tr}. The analysis also quantifies the influence of target \ac{RCS} on the mean capacity, a critical metric for \ac{ISAC} system design. The simulation framework, governed by the parameters enumerated in Table \ref{sim_params}, models a \(4\times 4\) \ac{MIMO} link employing \ac{OFDM} signaling with \(N = 256\) subcarriers. The analysis assumes $\mu^2 = 1$.
\begin{figure*}[tb]
    \centering
    \includegraphics[trim={18mm 1mm 1mm 0mm},clip,width=0.9\textwidth]{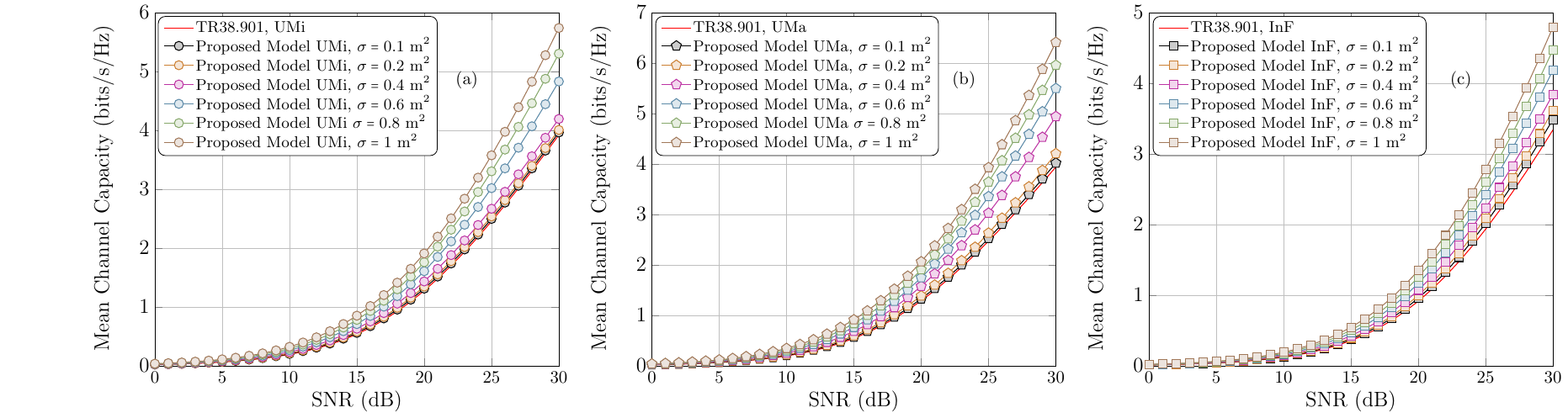} 
    \caption{Mean channel capacity comparison between the proposed \ac{ISAC} model and the \ac{TR}38.901 \cite{tr} baseline across a range of target \ac{RCS} values for: (a) \ac{UMi}, (b) \ac{UMa}, and (c) \ac{InF} scenarios.}
    \label{chan_cap} 
\end{figure*}

Fig. \ref{chan_cap} presents a comparative analysis of the mean channel capacity considering the proposed \ac{ISAC} channel model and the standard \ac{TR}38.901 \cite{tr} channel model for \ac{UMi}, \ac{UMa}, and \ac{InF} scenarios, respectively, evaluated across target \ac{RCS} values of \(0.1\)m\(^2\), \(0.2\)m\(^2\), \(0.4\)m\(^2\), \(0.6\)m\(^2\), \(0.8\)m\(^2\), and \(1\)m\(^2\). The results indicate that for lower \ac{RCS} values, e.g., \(0.1\)m\(^2\) and \(0.2\)m\(^2\), the mean channel capacity across all environments remains nearly identical to the \ac{TR}38.901 \cite{tr} baseline, as the minimal  path gain (where the received power is proportional to the target's \ac{RCS}) results in a negligible contribution to the composite \ac{CIR}. This small path gain is unable to generate significant additional eigenmodes of the environmental channel matrix, resulting in a minimal increase in the mean channel capacity. Conversely, a high \ac{RCS} target introduces strong specular reflections that significantly alter the composite \ac{CIR} resulting in strengthening of dominant eigenmodes, increasing the effective channel rank or increasing the multiplexing gain, thereby directly elevating the mean channel capacity beyond the \ac{TR}38.901 \cite{tr} baseline.
\subsection{Sensing Performance Evaluation}\label{sec4b}
This section presents performance evaluation of the sensing functionality utilizing the proposed \ac{ISAC} channel model. The analysis is conducted through a series of Monte Carlo simulations across standardized propagation environments, including \ac{UMa}, \ac{UMi}, and \ac{InF} scenarios. The sensing metrics we consider are target range estimation accuracy and detection performance. The latter is quantified by generating \ac{ROC} curves, which depict the critical trade-off between the probability of detection (\(P_\mathrm{d}\)) and the probability of false alarm (\(P_\mathrm{fa}\)) in different environments.
\begin{figure}[tb]
    \centering
    \includegraphics[trim={18mm 1mm 1mm 0mm},clip,width=0.45\textwidth]{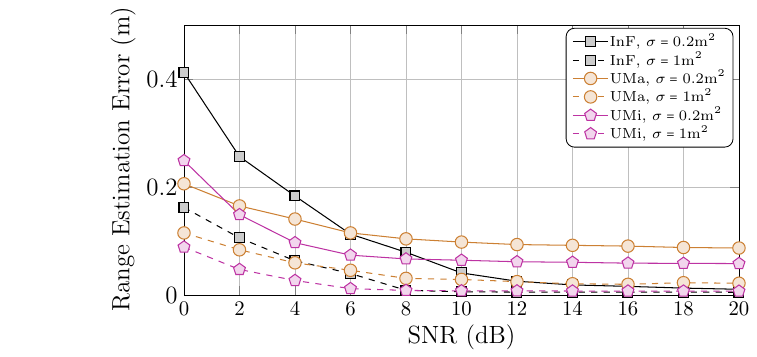} 
    \caption{Average target range estimation analysis as a function of \ac{SNR} for two targets with \acp{RCS} of \SI{0.2}{\meter^2} and \SI{1}{\meter^2} across \ac{UMa}, \ac{UMi}, and \ac{InF}  scenarios.}
    \label{pos_error} 
\end{figure}
\begin{figure*}[tb]
    \centering
   \includegraphics[trim={60mm 100mm 60mm 100mm},clip,width=0.9\textwidth]{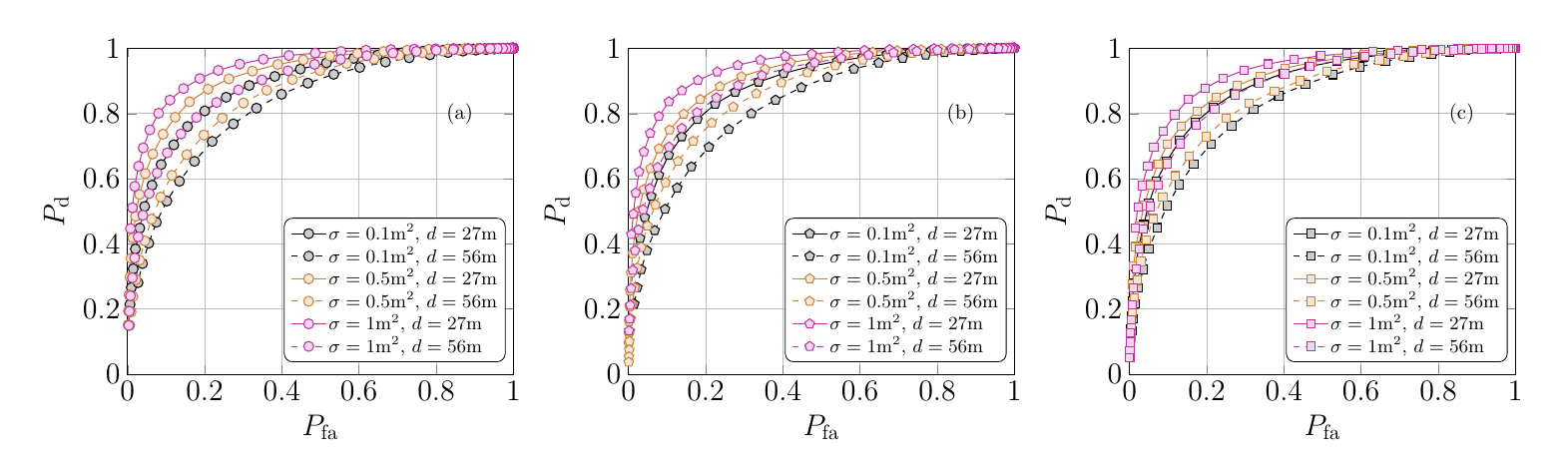} 
    \caption{\ac{ROC} showing \(P_\mathrm{d}\) versus \(P_\mathrm{fa}\) for (a) \ac{UMa}, (b) \ac{UMi}, and (c) \ac{InF} environments. Each subplot evaluates \ac{Tx}-target-\ac{Rx} distances of \SI{27}{\meter} and \SI{56}{\meter} across target \ac{RCS} values of \SI{0.1}{\meter^2}, \SI{0.5}{\meter^2}, and \SI{1}{\meter^2}.}
    \vspace{-0.7cm}
    \label{rocs} 
\end{figure*}

\subsubsection{Target Range Estimation}
This section evaluates target range estimation performance using the PARAMING algorithm \cite{salmane2025high} within the proposed \ac{ISAC} channel modeling framework. Simulations are carried out for the standardized \ac{UMa}, \ac{UMi}, and \ac{InF} deployment scenarios using the parameter set listed in Table \ref{sim_params}. Two sensing targets are considered: target 1, representing a small-sized UAV with an \ac{RCS} of \SI{0.2}{\meter^2}, and target 2, representing a human with an \ac{RCS} of \SI{1}{\meter^2}. Based on the \ac{Tx} and \ac{Rx} heights, this geometry is representative of a roadside-unit to roadside-unit sensing configuration. The Cartesian coordinates of both targets are provided in Table \ref{sim_params}. From these coordinates, it can be observed that target 1 has a shorter overall bistatic distance than target 2, but a smaller \ac{RCS}, whereas target 2 is located at a larger bistatic distance and exhibits a higher \ac{RCS}.

The range estimation results are shown in Fig. \ref{pos_error} which illustrates that reliable sensing is achievable in all considered environments provided that the \ac{SNR} is sufficiently high. A key observation is that increasing the target \ac{RCS} significantly improves localization performance, even when the bistatic distance is larger. In particular, target~2, despite being farther from the sensing nodes, can still be estimated with smaller positioning errors because of its stronger echo. At the same time, the results also show that smaller and closer targets, such as target 1, remain localizable, although with a somewhat higher estimation error due to their weaker scattering response.

Across the considered channel scenarios, \ac{UMa} consistently exhibits slightly worse performance for both targets, which can be attributed to its less favorable propagation conditions, including longer propagation distances, higher path loss, and richer multipath effects. In contrast, \ac{UMi} generally outperforms \ac{UMa} because of its stronger dominant \ac{LoS} components and reduced attenuation. The \ac{InF} and \ac{UMi} scenarios provide almost identical performance for target~2 with higher \ac{RCS}, indicating that for strong targets the dominant echo can be estimated reliably in both environments. For target~1, however, \ac{InF} achieves the best performance. This gain is mainly due to the propagation characteristics of the adopted \ac{InF} channel model, where the assumed clutter density of \(10\%\) reduces the contribution of secondary scatterers, mitigates multipath-induced ambiguities, and improves the extraction of the dominant target echo used for range estimation.
\subsubsection{Detection Performance via ROC Curves}
The following analysis evaluates the detection performance by quantifying the \(P_\mathrm{d}\) and \(P_\mathrm{fa}\). The evaluation is conducted for two distinct \ac{Tx}-target-\ac{Rx} distances, \(d = 27\)m and \(d = 56\)m, within standardized \ac{UMa}, \ac{UMi}, and \ac{InF} propagation environments. When \(d=27\)m, the Cartesian coordinates for the \ac{Tx}, target, and \ac{Rx} are \((0,0,10)\)m, \((0,5,10)\)m and \((5,15,10)\)m, respectively. On the other hand, for \(d=56\)m, the \ac{Tx} and \ac{Rx} coordinates remain the same, however, the target is \((25,15,10)\)m. Moreover, the target \ac{RCS} we consider are \(0.1\)m\(^2\), \(0.5\)m\(^2\) and \(1\)m\(^2\). The \ac{RCS} of \(0.1\)m\(^2\) is characteristic of a small-sized \ac{UAV}, whereas, the \ac{RCS} of a human adult is approximately \(1\)m\(^2\). All remaining system parameters are held constant, as specified in Table \ref{sim_params}. The detection mechanism employs an energy detector \cite{10681445}, which operates on the principle of binary hypothesis testing. 

The detection performance, quantified by \ac{ROC} curves across \ac{UMa}, \ac{UMi}, and \ac{InF} propagation environments, is illustrated in Fig. \ref{rocs}. The results demonstrates an expected dependence on target parameters, i.e., larger \ac{RCS} values and smaller \ac{Tx}-target-\ac{Rx} distances correspond to enhanced detection performance. This improvement is manifested as a higher \(P_\mathrm{d}\) for any given \(P_\mathrm{fa}\). Conversely, diminished \ac{RCS} or increased link range reduces the detection probability. The results confirm that target parameters significantly influence sensing reliability in different scenarios. From Fig. \ref{rocs}, we can observe that for \(P_\mathrm{d} = 0.9\), the \(P_\mathrm{fa}\) in \ac{UMa}, \ac{UMi}, and \ac{InF} is \(0.42\), \(0.4\) and \(0.44\) resp., when $\sigma= 0.5\text{m}^2$ and \(d = 56\)m.
\vspace{-0.3cm}
\section{Conclusions}\label{sec5}
We presented a channel model framework for  \ac{GBSM} bistatic \ac{ISAC} system. The core contributions are threefold: (1) the introduction of a dual-component \ac{ISAC} \ac{GBSM} that explicitly segregates target and background components, enabling concurrent communication and sensing performance evaluation, (2) the development of a hybrid clustering methodology that integrates geometrically deterministic clusters with traditional stochastic ones, enforcing the spatiotemporal consistency mandatory for sensing requirements, and (3) a rigorous cross-validation demonstrating that the proposed model retains full backward compatibility with \ac{TR}38.901 \cite{tr} for communication metrics (\ac{BER}, ergodic capacity) while providing the necessary fidelity to evaluate key sensing performance indicators. The proposed dual-component \ac{ISAC} \ac{GBSM} provides an essential foundation for the design, simulation, and validation for future \ac{ISAC} systems, offering a tool for analyzing the complex interplay and mutual constraints between \ac{SnC} functionalities in shared channels using an \ac{ISAC} channel simulator. Future work will extend the proposed \ac{ISAC} channel modeling framework by incorporating cluster birth-death processes and by characterizing channel non-stationarity under target motion with high velocity and arbitrary trajectories. A key open problem enabled by the proposed framework is the joint identification of target-induced and background multipath components from measurement data. In this context, the development of practical and reliable algorithms for labeling resolvable propagation paths as either target-related or background-related represents an important research direction. Another promising avenue is to leverage the proposed target/background channel decomposition for downstream sensing and inference tasks, such as target classification and human activity recognition, by combining target-channel isolation with feature extraction based on the temporal evolution of scattering points and their associated delay-angle-Doppler signatures.

\vspace{-0.3cm}
\bibliographystyle{IEEEtran}
\bibliography{biblio}

\end{document}